\documentclass[a4paper,11pt]{article}

\usepackage[utf8]{inputenc} 
\usepackage[T1]{fontenc}

\usepackage{latexsym}
\usepackage{ntheorem}

\usepackage{xcolor}
\usepackage{jheppub} 

\usepackage[T1]{fontenc} 
\definecolor{darkgreen}{rgb}{0.0, 0.5, 0.0}

\usepackage{url}

\usepackage{amssymb} 
\usepackage{xcolor} 
\usepackage{graphicx} 
\usepackage{caption}
\usepackage[labelformat = simple]{subcaption}

\usepackage[normalem]{ulem}
\usepackage{bbm}
\usepackage{dsfont}
\usepackage[mathscr]{euscript}
\usepackage{ntheorem}

\input{LambdaMacros}

\renewcommand{\redca}[1]{\color{darkgreen}\mnote{green text added#1}}

\usepackage{cancel}

\newcommand{\N}{\mathbb{N}}

\newcommand{\snD}{\cancel\nabla}

\newcommand{\smet}{\blue{g}}
\newcommand{\IMuu}{\alpha}
\def\ben{\begin{equation}}
\def\een{\end{equation}}
\def\bena{\begin{eqnarray}}
\def\eena{\end{eqnarray}}
\def\half{{1\over 2}}
\def\quater{{1 \over 4}}
\newcommand{\non}{\nonumber} 

\renewcommand{\ptcheck}[1]{}

 \author{Piotr T.\ Chru\'sciel}
 \author[1]{and Wan Cong \note{Corresponding author.}}
 \affiliation{University of Vienna, Faculty of Physics
  \\Boltzmanngasse 5, A 1090 Vienna, Austria}
 \emailAdd{piotr.chrusciel@univie.ac.at}
   \emailAdd{wan.cong@univie.ac.at}
\title{\boldmath Gluing variations
}

\abstract{
We establish several results on gluing/embedding/extending geometric structures in vacuum spacetimes with a cosmological constant in any spacetime dimensions $d\ge 4$, with emphasis on characteristic data. A useful tool is provided by the notion  of submanifold-data of order $k$. As an application of our methods we prove that vacuum Cauchy data on a spacelike Cauchy surface with boundary can always be extended to vacuum data defined beyond the boundary.
}

\renewcommand{\blue}[1]{{#1}}
\renewcommand{\red}[1]{{#1}}

\renewcommand{\hyp}{\red{\Sigma}}
\renewcommand{\refq}{\mnote{blue text added or changed following a query from a referee}\color{blue}}

\renewcommand{\redca}[1]{\color{red}\ptcr{change or addition or rewording on #1}}

\begin{document}
\maketitle
\flushbottom
\section{Introduction}

In a recent series of pioneering papers, Aretakis, Czimek and Rodnianski~\cite{ACR1,ACR2,ACR3} presented a gluing construction  for characteristic initial data for four-dimensional vacuum Einstein equations. The purpose of this paper is to show that \emph{related} gluing constructions can be done using a
spacelike gluing \emph{\`a la} Corvino~\cite{Corvino}.
While the construction in~\cite{ACR1,ACR2,ACR3} uses the structure of the four-dimensional Einstein equations in a substantial way, our approach applies to any dimensions. As a bonus, we allow  a non-vanishing cosmological constant. The resulting spacetimes are essentially identical, but the intermediate steps are different.

As such, the general relativistic  gluing problem can be viewed as the following question: given two spacetimes, solutions of vacuum Einstein equations, can one find a third one where non-trivial subsets of each of the original spacetimes are isometrically included?

A version of this can be formulated at the level of spacelike Cauchy data: Consider a manifold $\hyp$ and two vacuum initial data sets $(\hyp_1,g_1,K_1)$ and $(\hyp_2,g_2,K_2)$ defined on overlapping subsets $\hyp_1$ and $\hyp_2$ of $\hyp$.
Can one find a vacuum data set $(g,K)$ on $\hyp$ which  coincides with the original ones away from the overlap, or away from a small neighborhood of the common boundary?
A positive answer to this has first been given by Corvino~\cite{Corvino} in a restricted setting, and  generalised in~\cite{CorvinoSchoen2,ChDelay}; see~\cite{ChBourbaki,CarlottoLR} for further references.
The problem is well understood for data sets which   are not-too-far-away from each other in the overlap: the gluing can be performed if the spacetimes $(\mcM_1,\fourg_1)$ and $(\mcM_2,\fourg_2)$, obtained by evolving  $(\hyp_1,g_1,K_1)$ and $(\hyp_2,g_2,K_2)$, have no Killing vectors near the overlapping region. Equivalently, the set of Killing Initial Data (KIDs) on the overlap is trivial.

\begin{figure}[h]
  \centering
  \begin{subfigure}[t]{0.8\textwidth}
  \centering
\includegraphics[scale=0.3]{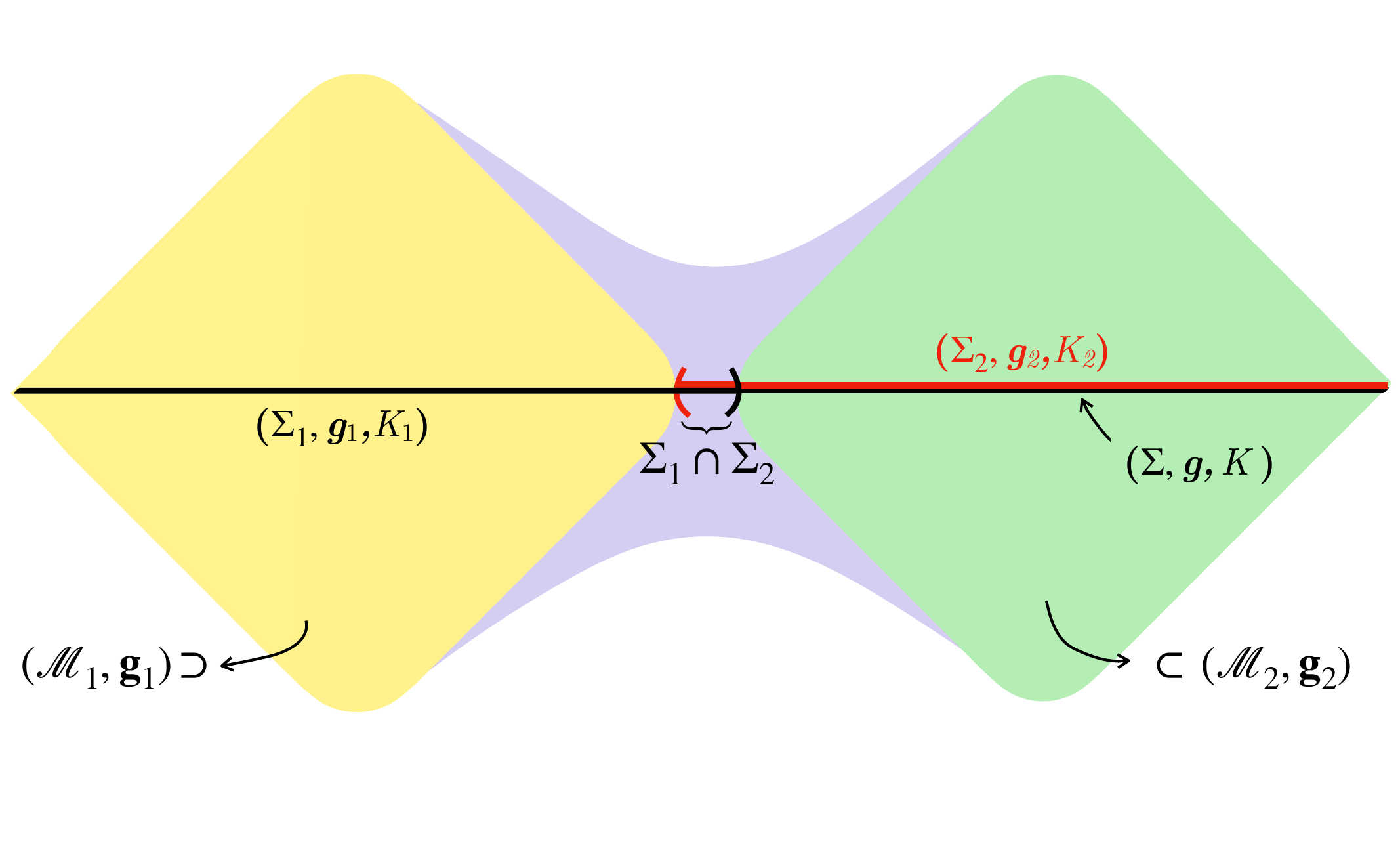}
\caption[a.]{Gluing overlapping spacelike initial data sets.
}
   \label{F17XII22.1a}
\end{subfigure}
\begin{subfigure}[t]{0.8\textwidth}
  \centering
\includegraphics[scale=0.3]{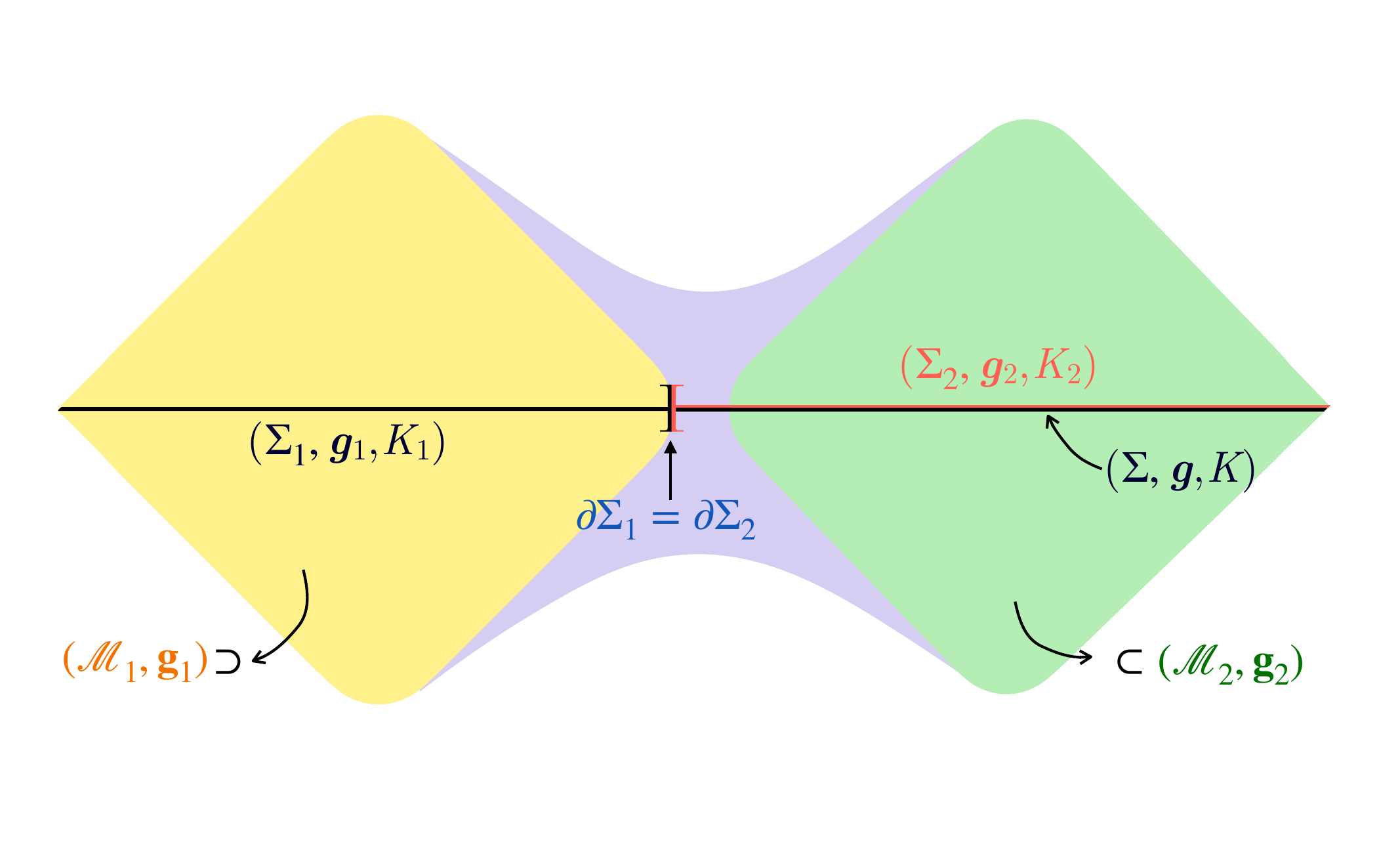}
\caption{Gluing touching spacelike initial data sets.
}
   \label{F17XII22.1b}
   \end{subfigure}
   \caption{Spacelike gluings. The metric is smooth and vacuum everywhere, identical to the original one in the left (yellow) and right (green) regions, the metric in the middle (violet) region interpolates smoothly between the original ones.}
\end{figure}

Note that a  gluing of overlapping spacelike initial data leads to a gluing of spacetimes in the following sense:  the domains of dependence of $\hyp_1\setminus \hyp_2$ and  $\hyp_2\setminus\hyp_1$, within the spacetime obtained by evolving the data on $\hyp$,  are isometric to the corresponding domains of dependence in the original spacetimes $(\mcM_1,\fourg_1)$ and $(\mcM_2,\fourg_2)$,  see Figure~\ref{F17XII22.1a}.

An essentially identical construction applies  with $\hyp_1$ and $\hyp_2$ lying on opposite sides of a common boundary, $\partial \hyp_1 = \partial \hyp_2$, see Figure~\ref{F17XII22.1b}.

One is then  led to the question, whether something similar can be done using  null  initial data.%
\footnote{
Classic works concerned with characteristic spacetime gluing include~\cite{BarrabesIsrael,Israel66,KhanPenrose}.
 While~\cite{BarrabesIsrael,Israel66} focus on lightlike shells, the point of our constructions is to avoid occurrence of such shells.
}
For instance, consider a
smooth hypersurface $\mcN $ and two characteristic data sets on overlapping subsets $\mcN_1$ and $\mcN_2$ of $\mcN$. Suppose that the data on both $\mcN_1\subset \mcM_1$ and $\mcN_2\subset \mcM_2$ arise by restriction from  vacuum spacetimes  $ (\mcM_1,\fourg_1)$  and $(\mcM_2,\fourg_2)$.  Can one find  a vacuum spacetime $(\mcM ,\fourg)$, with $\mcN\subset \mcM$, so that the data on $\mcN $, arising by restriction from $\fourg$, coincide with the original ones away from the overlapping region?

Here the situation is somewhat different, as a well-posed characteristic
initial-value problem requires either two \emph{transverse} initial-data surfaces $\mcNone$ and $\mcNtwo$ (not to be confused with the hypersurfaces $\mcN_1$ and $\mcN_2$ considered above \red{and in what follows},
which are included in a single smooth hypersurface $\mcN$), or a light cone.
This makes it
clear that an answer in terms of characteristic initial data on
a single smooth hypersurface is not possible. However, given $k\in \N$, one can complement the characteristic initial data on $\mcN_1$ and $\mcN_2$ with information about $k$  derivatives of the metric in directions transverse to $\mcN_1$ and $\mcN_2$;  such transverse derivatives can be obtained by solving transport equations (i.e., ODE's along the generators) on $\mcN_1$ and $\mcN_2$ from  data on   cross-sections $\secN_1\subset \mcM_1$ and $\secN_2\subset \mcM_2$ after some gauge choices have been made.
Denoting by $\CSdata{\mcN_1,k}$ a set of characteristic data on $\mcN_1$ together with transverse derivatives up to order $k\in \N\cup\{\infty\}$, and by $\CSdata{\mcN_2,k}$  the set of such data on $\mcN_2$, one can ask whether there exist data $\CSdata{\mcN,k}$ on $\mcN$ which coincide with the original data on the overlap region.

Explicit parameterisations of the data $\CSdata{\mcN ,k}$ are presented in Sections~\ref{s6I23.1}, \ref{s22XII22.2}, and in Appendix~\ref{app22XII22.1}. The question of optimal differentiability conditions of the fields parameterising  $\CSdata{\mcN ,k}$ is delicate, and for simplicity
we will require the existence of local coordinate systems near $\mcN$ in which all the functions parameterising $\CSdata{\mcN ,k}$ are smooth  \emph{on $\mcN$}.

\begin{figure}[h]
  \centering
\includegraphics[scale=0.3]{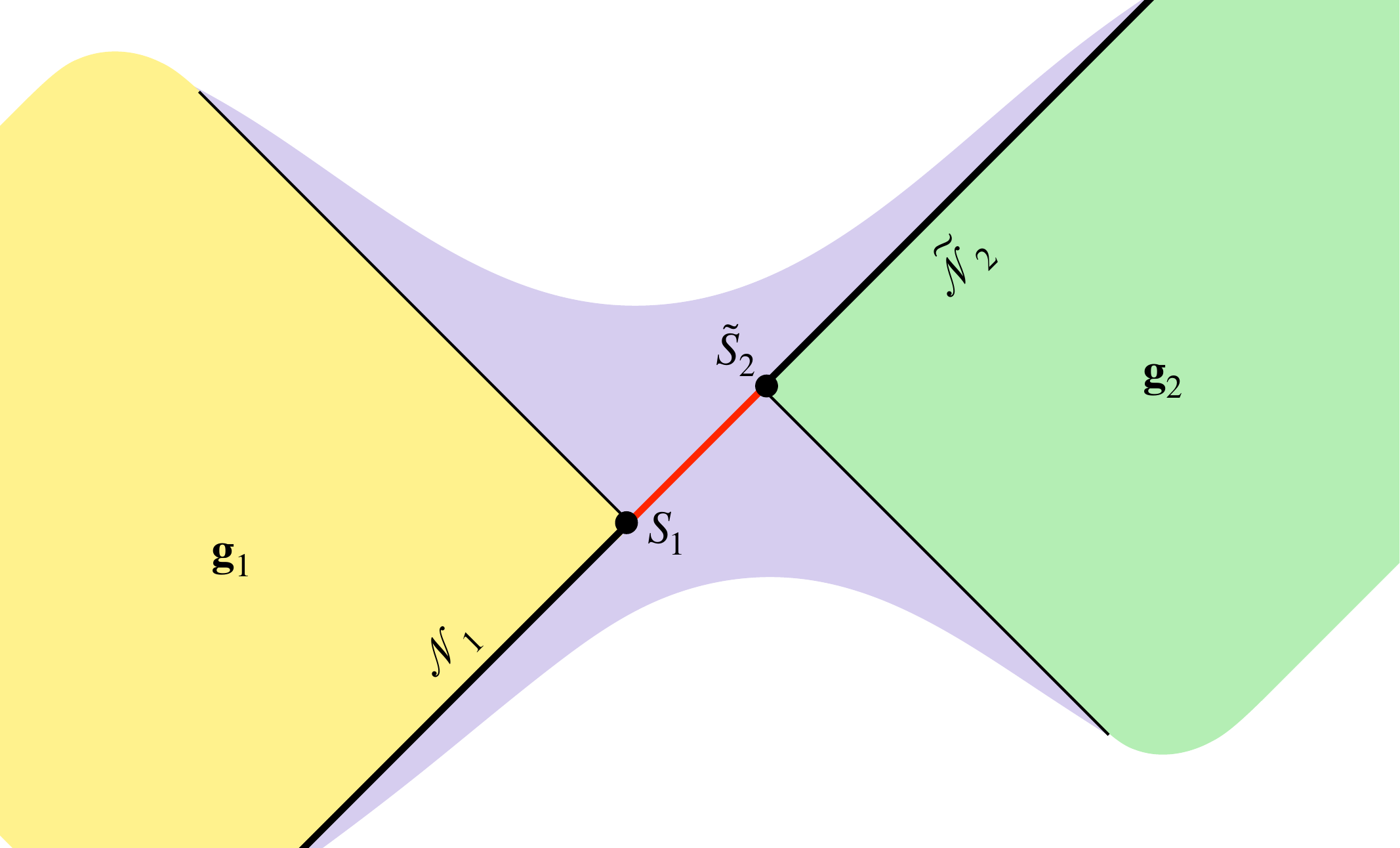}
   \caption{Gluing construction of \cite{ACR1}.}
   \label{F13II23.1}
\end{figure}

In their landmark work, Aretakis, Czimek and Rodnianski have given a positive answer to the following variation of the gluing question, in a near-Minkowskian setting, illustrated in Figure~\ref{F13II23.1}. Namely,  supposing that $\CSdata{\mcN_1,k}$ are close to $\CSdata{\mcN_2,k}$ on the overlap region, one asks:

\begin{question}
  \label{Q17XII22.1}
  Do there exist characteristic data $\CSdata{\tmcN2,k}$ on  a null hypersurface
  $ \tmcN2$, obtained by slightly moving $\mcN_2$ in $(\mcM_2,\fourg_2)$,  and characteristic data $\CSdata{\tmcN{},k}$, on a null hypersurface $\tmcN{}$ connecting
   $\mcN_1\setminus \mcN_2$ and $\tmcN2$, which coincide with $\CSdata{\mcN_1\setminus \mcN_2,k}$ on $\mcN_1\setminus \mcN_2$ and with the data $\CSdata{\tmcN2,k}$ on $\tmcN2$?
\end{question}

  The authors of~\cite{ACR1,ACR2,ACR3} assume that $\mcN$ has the topology of a light cone in four-dimensional Minkowski spacetime and that $k=2$.  The overlap   is taken to be far away from the tip of the light cone, so that regularity issues at the tip are irrelevant.
  They show that  there exists a ten-parameter family of obstructions to the gluing.
In the case where the data $\CSdata{\mcN_2,2}$ arise from a null hypersurface in a Kerr space-time, they show that one can get rid of the obstructions by adjusting the mass and angular momentum of the Kerr metric. In~\cite{CzimekRodnianski} a striking extension of the method is presented, where the obstructions are reduced to a single one, namely a lower bound on the mass of the Kerr extension.

In~\cite{ACR3} it is also shown how to make a Corvino-type spacelike gluing using the characteristic gluing.
%

The aim of this work is to show
that spacelike gluings can be used to construct spacetimes with properties similar to those resulting from the construction of  \cite{ACR1}. In the approach described here the hypersurface $\tmcN2$ of Question~\ref{Q17XII22.1} is again obtained by moving slightly $\mcN_2$ within $(\mcM_2,\fourg_2)$, but we give up the requirement that   $\mcN_1$ and $\tmcN2$ are subsets of the same smooth null hypersurface in the final spacetime. Indeed, in our construction the null hypersurface extending smoothly $\mcN_1$ in the new spacetime is obtained by  moving $\tmcN2$
in an auxiliary, suitably constructed, nearby vacuum metric; see Figure~\ref{F17XII22.2c} below. Our method does \emph{not} provide a \emph{null gluing}, but a variation thereof; hence the title of this paper.

A useful tool in this context is provided by \emph{submanifold data of order $k$}, introduced in Section~\ref{s8I23.11}. We provide a simple parameterisation of vacuum characteristic data on null hypersurfaces and on submanifolds of codimension-two in Section~\ref{s6I23.1} using coordinate systems introduced by Isenberg and Moncrief~\cite{VinceJimcompactCauchyCMP}.
A second such parameterisation is provided in Section~\ref{s22XII22.2}  using Bondi coordinates. In Section~\ref{s26XII22.1} we use a ``hand-crank construction'' to show that vacuum characteristic initial data, or vacuum spacelike data on a submanifold of codimension 2, of any order can be realised by embedding as an \emph{interior submanifold} of a smooth vacuum spacetime.
As Corollaries we obtain that any spacelike general relativistic vacuum Cauchy data on a manifold with smooth boundary can be extended to a larger vacuum initial data set,
 where the original boundary becomes an interior submanifold (cf. Theorem~\ref{T10III23.1} below; compare \cite{SmithWeinstein1,SmithWeinstein2,MantoulidisSchoen,Czimek:2016ydb,Bartnik:quasi-sph,CederbaumExtensions} for similar results under restrictive conditions), and that any vacuum data $\CSdata{\{p\},k}$ at a point $p$ can be realised by a vacuum metric (cf.\ Corollary~\ref{C4II23.2}).
In Section~\ref{s22XII22.1}
 we use a ``Fledermaus  construction'' to show that vacuum characteristic initial data on two transversely intersecting hypersurfaces can likewise be realised by embedding as an \emph{interior submanifold with corner} of a smooth vacuum spacetime.
In Section~\ref{s26XII22.3} we show how to carry out a variation of the characteristic gluing using spacelike gluing.
We apply this result in Section~\ref{s4II23.1} to glue two sets of cross-section data, one of them arising from the Kerr family.
  In Appendix~\ref{app22XII22.1} we show how the sphere data of \cite{ACR1} relate to our codimension-two data in Bondi parameterisation. In Appendix~\ref{s18IV22.1asdf} we show existence of a preferred, unique set of Bondi coordinates associated with a null hypersurface $\mcN$ with a cross-section $\secN$. In Appendix~\ref{CWs23II22.1}
we show, by quite general considerations, that spacetime Killing vectors provide an obstruction to  certain gluing constructions;
see Remark~\ref{R22I23.2} and Section~\ref{s4II23.1} below for further comments on this.

Throughout this work ``vacuum'' means a solution of the vacuum Einstein equations with a cosmological constant $\Lambda\in \R$ in spacetime dimension $n+1\ge 3+1$.

\section{The existence theorem for two null hypersurfaces intersecting transversally}
 \label{s28XII22.1}

In what follows we will need an existence theorem for the characteristic Cauchy problem, and the aim of this section is to review a version thereof. This allows us also to introduce our notations.

 Let $\mcN$ be a smooth null hypersurface in an $(n+1)$-dimensional spacetime $ (\mcM,\fourg)$. Introduce a coordinate system $(u,r,x^A)$ in which $\mcN=\{u=0\}$ and in which $\partial_r$ is tangent to the null geodesics threading $\mcN$, and with $\fourg (\partial_r,\partial_{x^A})|_{\mcN} = 0$.

  In this section, for consistency with~\cite{ChPaetz} we write
  $\og _{\mu\nu}$ to denote  the restriction of  $\fourg _{\mu\nu}$ to $\mcN$, also referred to as the \emph{trace of $\fourg _{\mu\nu}$ on $\mcN$}.%
  \footnote{
  However, we will not do this in the remaining sections, hoping that the domains of definition will be clear from the context. Nevertheless, we will consistently 
 use the boldface symbol $\fourg$ for the spacetime metric  defined on open subset of spacetime, and $g$ (or $\overline g$ in this section) for the metric restricted or induced on submanifolds.
}
  The trace should not be confused with  pull-backs: the pull-back to $\{u=0\}$ of a tensor field $A_{uu} du^2$ is zero, while the trace of $A_{uu}du^2$ to $\{u=0\}$
  is the tensor field $A_{uu}|_{u=0}du^2$ defined along $\{u=0\}$, which vanishes if and only if $A_{uu}|_{u=0} \equiv 0$.

On $\mcN $ let
\begin{equation}
\thetatau \equiv \frac{1}{2}\ovg ^{AB}\partial_r \ovg _{AB}
\label{dfn_tau}
\end{equation}
be the divergence scalar,
and let
\begin{equation}
 \sigma_{AB} \equiv \frac{1}{2} \partial_r \ovg _{AB} - \frac{1}{n-1}\thetatau \ovg _{AB}
\label{dfn_sigma}
\end{equation}
 be the trace-free part of $\frac 12 \partial_ r  \ovg _{AB}$, also known as the \emph{shear tensor}.
 The vacuum Raychaudhuri equation,
\begin{equation}
\partial_r\thetatau - \kappa \thetatau   + |\sigma|^2 + \frac{\thetatau^2}{n-1} = 0
 \,,
  \label{21XII11.30x}
\end{equation}
where
\begin{equation}
 |\sigma|^2 \equiv \sigma_A{}^B \sigma_B{}^A \,,
 \quad \sigma_A{}^B \equiv \ovg ^{BC} \sigma_{AC}
 \,,
\end{equation}
and
\begin{equation}\label{4I23.11}
  \kappa =  {\Gamma}{^r_{rr}}|_{u=0}
\end{equation}
(see \cite[Appendix~A]{CCM2} for a collection of explicit formulae in adapted coordinates) provides on $\mcN$ a constraint equation for the family of $(n-1)$-dimensional metrics
$$
 r\mapsto \ovg _{AB}(r,x^C)\mathrm{d}x^A\mathrm{d}x^ B
 \,.
$$
The geometric meaning of $\kappa$ is that of the connection coefficient of the one-dimensional bundle of tangents to the null generators of $\mcN$, viewed as a bundle along each of the generators.
Indeed, under a change of coordinates $(r,x^A)\mapsto (\mybar r(r,x^A), \mybar x^A = x^A)$ we have
\begin{equation}\label{15I23.5}
 \tautheta\mapsto  \mybar \tautheta
   := \frac{1}{2}\ovg ^{AB}\partial_{\mybar r } \ovg _{AB}
   = \frac{\partial r}{\partial \mybar r} \tautheta
   \,,
   \quad
   \sigma_{AB}\mapsto
\mybar  \sigma_{AB} :=  \frac{1}{2} \partial_{\mybar r} \ovg _{AB} - \frac{1}{n-1}\mybar \thetatau \ovg _{AB}=
 \frac{\partial r}{\partial \mybar r}
  \sigma_{AB}
 \,,
\end{equation}
and  the Raychaudhuri equation becomes
\begin{equation}
\partial_{\mybar r} \mybar \thetatau - \mybar \kappa \mybar \thetatau   + |\mybar \sigma|^2 + \frac{\mybar\thetatau^2}{n-1} = 0
 \,,
  \label{15I23.6}
\end{equation}
with
\begin{equation}
\kappa \mapsto \mybar \kappa
  = \frac{\partial r}{\partial \mybar r}  \kappa
  +
  \frac{\partial \mybar r}{\partial  r}
  \frac{\partial^2 r}{\partial \mybar r ^2}
  \,.
  \label{15I23.7}
\end{equation}
 The vanishing of $\kappa$ is equivalent to the requirement that $r$ is an affine parameter along the generator.

We are ready now to pass to the problem at hand.
Consider two smooth hypersurfaces $\mcNone $ and $\mcNtwo$ in an $(n+1)$--dimensional manifold $\mcM $, with  transverse intersection at a smooth submanifold $\secN$.
We choose adapted coordinates $(\xone ,\xtwo ,x^A)$ so that    $\mcNone $  coincides with the set $\{\xone =0\}$, while $\mcNtwo  $ is given  by $\{\xtwo =0\}$.
We suppose that $\mcNone = \Ione \times \secN $ and $\mcNtwo  = \Itwo \times \secN $, where $\Ione$ and $\Itwo$ are intervals of the form $ [0,\eendxone )$   for some $\eendxone  \in \R^+\cup\{\infty\}$, possibly with distinct $\eendxone$, with the coordinate $\xone$ ranging over $\Itwo$  and $\xtwo$ ranging over $\Ione$.

We wish for the restriction  $\ovg _{\mu\nu}$  of the metric functions $\fourg _{\mu\nu}$ to the initial surface $\mcNone \cup \mcNtwo  $   to arise from a smooth Lorentzian spacetime metric.
This will be the case if $\ovg _{\xone\xtwo} $ is nowhere vanishing, if  $\xtwo \mapsto \ovg _{AB}(\xtwo,\cdot)$ is a family of Riemannian metrics on $\{\xtwo\}\times \secN\subset\mcNone$, similarly for  $\xone \mapsto \ovg _{AB}(\xone,\cdot)$ on $\{\xone\}\times \secN\subset\mcNtwo$, and if $\ovg _{\mu\nu}$  is
smooth on $\mcNone $ and $\mcNtwo  $ and continuous across $S\equiv \mcNone \cap \mcNtwo  $.
The hypothesis that   $\mcNone$ and $\mcNtwo$ are characteristic translates to
\begin{equation}\label{2I23.1}
  \fourg_{\xone\xone} \big|_{\mcNtwo  } \equiv 0\equiv \fourg_{\xone A} \big|_{\mcNtwo  }
  \,,
  \qquad
  \fourg_{\xtwo\,\xtwo} \big|_{\mcNone } \equiv0\equiv \fourg_{\xtwo A} \big|_{\mcNone }
  \,.
\end{equation}
We therefore  impose the following conditions on $\secN$:
%
\begin{eqnarray}
 &
   \displaystyle
  \lim_{\xtwo \rightarrow 0} \fourg_{AB}|_{\mcNone }
   \displaystyle
   = \
   \lim_{\xone \rightarrow 0} \fourg_{AB}|_{\mcNtwo  } \,,&
 \label{regularity1a}
\\
&
   \displaystyle
  \lim_{\xtwo \rightarrow 0} \fourg_{\xone\xtwo} |_{\mcNone }
   \displaystyle
   = \
   \lim_{\xone \rightarrow 0} \fourg_{\xone\xtwo} |_{\mcNtwo  }  \,,&
 \label{regularity1b}
\\
&
   \displaystyle
  \lim_{\xtwo \rightarrow 0} \fourg_{\xone A} |_{\mcNone }
   = \    0  \,, \quad   \lim_{\xone \rightarrow 0} \fourg_{\xtwo A} |_{\mcNtwo  } =\, 0
 \,,
 &
 \label{regularity1c}
\\
&
   \displaystyle
 \lim_{\xtwo \rightarrow 0} \fourg_{\xone\xone} |_{\mcNone }
   \displaystyle
   = \  0  \,, \quad  \lim_{\xone \rightarrow 0} \fourg_{\xtwo\,\xtwo} |_{\mcNtwo  }=\, 0
 \,.
 &
 \label{regularity1d}
\end{eqnarray}

To avoid ambiguities, we will denote by $\kappa _{\mcN }$ the function $\kappa$ associated with the hypersurface $\mcN $; similarly for  $\kappa _{\mcNtwo }$, $\thetatau_{\mcNone} $,  etc.
We choose the time-orientation by requiring that both $\partial_\xone$ and $\partial_\xtwo$ are  future-oriented at $\secN$.
 We have the following~\cite{ChPaetz}:%
\footnote{We take this opportunity to point out two annoying misprints in \cite{ChPaetz}, corrected here: in Equation (40a) there the left-hand side should be $\fourg |_{N_{2}}$, and in Equation (40b) the left-hand side should be $\fourg |_{N_{1}}$.}

\begin{theorem}
 \label{T5I23.1}
Let there be given functions $(\ovg _{\mu\nu},\kappa)$ on $\mcNone \cup \mcNtwo  $,
smooth up-to-boundary on $\mcNone $ and $\mcNtwo  $, such that
%
\begin{eqnarray}
 \label{29X0.1a2}
 \fourg_{\mu\nu}|_{\mcNone  } \mathrm{d}x^\mu \mathrm{d}x^\nu =\ovg _{\xone\xone} (\mathrm{d}\xone )^{2}+2\ovg _{\xone\xtwo} \mathrm{d}\xone \, \mathrm{d}\xtwo
 +2\ovg _{\xone A} \mathrm{d}\xone \mathrm{d}x^{A}+
 \ovg _{AB}\mathrm{d}x^{A}\mathrm{d}x^{B}
 \,,
\\
 \label{29X0.1a1}
 \fourg_{\mu\nu}|_{\mcNtwo  } \mathrm{d}x^\mu \mathrm{d}x^\nu =\ovg _{\xtwo\,\xtwo} (\mathrm{d}\xtwo )^{2}+2\ovg _{\xone\xtwo} \mathrm{d}\xone \, \mathrm{d}\xtwo
 +2\ovg _{\xtwo A} \mathrm{d}\xtwo \mathrm{d}x^{A}+
 \ovg _{AB}\mathrm{d}x^{A}\mathrm{d}x^{B}
 \,,
\end{eqnarray}
%
{
 with $\mcNone=\{\xone=0\}$ and $\mcNtwo=\{\xtwo=0\}$.
 }
Suppose that
$\ovg _{\xone\xtwo}<0$, that \eqref{regularity1a}-\eqref{regularity1d} hold, and that
%
\begin{eqnarray}
 &
 \label{21XII11.2}
 \partial_{\xone} \ovg_{\xtwo\,\xtwo} |_\secN =    2\big(   \partial_{\xtwo} \ovg_{\xone\xtwo}  -\ovg _{\xone\xtwo}  \kappa_{\mcNone }  \big)|_\secN
  \qquad
 \Big(
  \Longleftrightarrow \quad \Gamma^\xtwo _{\xtwo\,\xtwo} |_\secN = \kappa_{\mcNone }|_\secN
  \Big)
  \,,
  &
\\
 &
 \label{21XII11.3}
 \partial_{\xtwo} \ovg_{\xone\xone} |_\secN =   2  \big( \partial_{\xone} \ovg_{\xone\xtwo}  -\ovg_{\xone\xtwo}  \kappa_{\mcNtwo  }  \big)|_\secN
   \qquad
 \Big(
 \Longleftrightarrow \quad   \Gamma^\xone _{\xone\xone} |_\secN = \kappa_{\mcNtwo  }|_\secN
 \Big)
  \,.
  &
\end{eqnarray}
If the Raychaudhuri equation  holds both on $\mcNone$ and $\mcNtwo$, i.e.,
%
 \begin{eqnarray}
  -\partial_{\xtwo}\thetatau_{\mcNone } +  { \kappa _{\mcNone }} \thetatau_{\mcNone } - |\sigma _{\mcNone }|^2  - \frac{\thetatau^2  _{\mcNone }}{n-1} = 0 
   \ \mbox{on $\mcNone $}
 \,,
  \label{21XII11.40}
 \\
  -\partial_{\xone}\thetatau_{\mcNtwo  } +  { \kappa _{\mcNtwo  }} \thetatau_{\mcNtwo  } - |\sigma _{\mcNtwo  }|^2  - \frac{\thetatau^2  _{\mcNtwo  }}{n-1} =0 
   \ \mbox{on $\mcNtwo  $}
 \,,
  \label{21XII11.41}
\end{eqnarray}
%
then
there exists a smooth metric defined on a neighbourhood of 
 solving the vacuum
 Einstein equations to the future of $\mcNone \cup \mcNtwo  $
  and realising the data  $(\ovg _{\mu\nu},\kappa)$ on $\mcNone \cup \mcNtwo  $.
\end{theorem}

\begin{Remark}\label{R14I23.2}
{\rm 
{
The existence of a solution in a one-sided neighborhood \emph{of $S$} has been established in the pioneering work of Rendall~\cite{RendallCIVP} in spacetime dimension $n+1=4$, and in \cite{CCM2} in all higher dimensions.
Our statement points out that a solution exists in a  one-sided neighborhood of $\mcNone\cup\mcNtwo$~\cite{Luk,CCW,RodnianskiShlapentokh,Collingbourne,%
ChruscielWafoGray}.
} 
}
\qed
\end{Remark}

\begin{Remark}\label{R14I23.6}
{\rm
The maximal globally hyperbolic solution is uniquely determined by the data, up to isometry. This can be seen by first noting that the local solution is defined uniquely, up to isometry, by the data listed; see~\cite{MarsCharacteristic2} for an extensive discussion.
Choosing a spacelike Cauchy hypersurface within the domain of existence of the coordinate solution, one can appeal to the usual Choquet-Bruhat -- Geroch uniqueness theorem for the spacelike Cauchy problem to conclude.
}
\qed
\end{Remark}

\begin{Remark}\label{R14I23.3}
{\rm
Rendall~\cite{RendallCIVP} requires that $\kappa\equiv 0$ and that the coordinates are solutions of the wave equation, $\Box_g x^\mu = 0$. The Raychaudhuri equations \eqref{21XII11.40}-\eqref{21XII11.41}
 are then solved using a conformal ansatz $  \og_{AB} =\phi^2 \gamma_{AB}$, with $\gamma_{AB}$ being freely prescribable on $\mcNone\cup\mcNtwo$ subject to continuity at $\secN$. These conditions determine the metric functions $\og_{\mu\nu}$ in terms of the free data uniquely up to the choice of
$$( \og_{\xone \xtwo}, \phi,\partial_\xone \phi,\partial_\xtwo \phi, \zeta_A)\big|_\secN
 \,,
$$
where $\zeta_A$ is the \emph{torsion 1-form} defined as follows: Assuming $\fourg _{\xone \xtwo}|_S$ is negative  we set $L=\sqrt{-2\ovg ^{\xone \xtwo}}\,\partial_\xtwo$ and
$ \genL =  \sqrt{-2 \og^{\xone \xtwo}}\, \partial_\xone$
along $\mcNone\cup\mcNtwo$, then $\zeta_A$ on $\secN$ is defined as
 \ptcheck{14I by wan} 
\begin{eqnarray}
 \zeta_A  & := &
 \frac{1}{2} \fourg(\nabla_A L, \underline L)\big|_\secN
 \equiv
 \frac{1}{2} \fourg^{\xone \xtwo}(\partial_\xone
 \fourg_{\xtwo A} - \partial_\xtwo  \fourg_{\xone A} )
 \big|_\secN
 \,.
\label{dfn_zeta_A}
\end{eqnarray}
Note that in our formulation of the characteristic initial value problem the torsion covector can be calculated from the remaining data in Theorem~\ref{T5I23.1}, as needed for Rendall's version of this theorem.
}
\qed
\end{Remark}
%
%
%

\section{Submanifold data}
 \label{s8I23.11}

Let $\mcM$ be an $(n+1)$-dimensional manifold.
Recall that, for $k\in \N\cup\{\infty\}$ and $p\in \mcM$, the $k$-th  jet $j^k_p A$  at $p$ of a tensor field  $A$, or of a function $A$, is the collection of all partial derivatives of $A$ up to order $k$ at $p$.
We also recall that this notion is coordinate-independent, in an obvious sense: given a representative of $j^k_p A$ in some coordinate system, we can determine $j^k_pA$ in any other coordinate system by standard calculus formulae.

Let $\subman $ be a smooth submanifold of $\mcM$.
For $k\in \N\cup\{\infty\}$ we define \emph{submanifold data of order $k$}  as the following collection of  jets \emph{defined along $\subman$}
\begin{equation}\label{8I23.7a}
  \CSdata{\subman,k} :=  \{ j^k_p\fourg _{\mu\nu}, p\in \subman\}
  \,,
\end{equation}
where we further assume that the jets arise by restriction to $\subman$ of jets of a smooth Lorentzian metric $\fourg _{\mu\nu}$ defined in a neighborhood of $\subman$.
Equivalently, $\CSdata{\subman,k}$ are those sections over $\subman$ of the bundle of jets which are obtained by calculating the jets of a spacetime metric defined near $\subman$ and restricting to $\subman$.
Or in yet equivalent words: an element of $\CSdata{\subman,k}$ is obtained by taking a Lorentzian metric defined in a neighborhood of $\subman$, calculating all its derivatives up to order $k$, and restricting the resulting fields to $\subman$.

We have assumed for simplicity that all the fields occurring in $\CSdata{\subman,k}$ are smooth, though one could of course consider more general situations.

It should be clear that an element of $ \CSdata{\subman,k} $ can always be extended to a smooth metric defined in a neighborhood of $\subman$ by using a {
 Taylor
expansion with a finite number of terms if $k$ is finite, or by Borel summation~\cite{BorelMem,Borel2}
}
if $k=\infty$, with the jets of the extended field inducing the original jets on $\subman$. Such an extension will be called \emph{compatible}.

Since a field  $\fourg _{\mu\nu}$ restricted to $\subman$ already carries information about its derivatives in directions tangent to $\subman$, for $k\ge 1$ the new information in  $\CSdata{\subman,k}$ is contained in the derivatives in directions transverse to $\subman$.

As a special case consider a  hypersurface $\hyp$. An equivalent definition of $\CSdata{\hyp,k}$ is obtained by choosing a smooth vector field  $X$   defined in a neighborhood of $\hyp$, transverse to $\hyp$, and setting
\begin{equation}\label{8I23.7b}
  \CSdata{\hyp,k}:= \Big(
   \big((\mcL_X)^i\fourg _{\mu\nu}\big)
   \big|_{\hyp}
  \Big)_{0\le i\leq k}
  \,.
\end{equation}
Given another smooth vector field $Y$ transverse to $\hyp$, the collection $
\Big(
   \big((\mcL_Y)^i\fourg _{\mu\nu}\big)
   \big|_{\hyp}
  \Big)_{0\le i\leq k}$ can be rewritten in terms of   $ \Big(
   \big((\mcL_X)^i\fourg _{\mu\nu}\big)
   \big|_{\hyp}
  \Big)_{0\le i\leq k}$, and vice-versa.

We will say that  $\CSdata{\subman,k}$ is spacelike if only metrics $\fourg _{\mu\nu}$ with spacelike pull-back to $\subman$ are allowed in \eqref{8I23.7a},
similarly for timelike, characteristic, etc.

The submanifold data $\CSdata{\subman,k}$ will be called \emph{vacuum} if the derivatives of the metric up to order $k$ satisfy the restrictions arising from the vacuum Einstein equations, including the equations obtained by differentiating the vacuum Einstein equations in transverse directions.
The structure of the set of vacuum jets is best understood by choosing preferred coordinates. For example, in normal coordinates at a point $p$ the first derivatives of the metric vanish and the second derivatives are uniquely determined by the Riemann tensor at $p$. In these coordinates the set of vacuum jets of second order is a subspace determined by a set of simple linear equations.

As another example, let $\hyp$ be again of codimension-one, and for $k\ge 2$ consider the set of vacuum data $\CSdata{\hyp,k}$ such that $\fourg _{\mu\nu}$ induces a Riemannian metric on $\hyp$. Then every element of $\CSdata{\hyp,k}$ can be equivalently described by the usual vacuum general relativistic initial data fields $(g,K)$, where $g $ is a Riemannian metric on $\hyp$, $K$ represents the extrinsic curvature tensor of $\hyp$, with $(g,K)$ satisfying the  (spacelike) constraint equations. Indeed, one can then introduce e.g.\ harmonic coordinates for $\fourg _{\mu\nu}$, and algebraically determine all transverse derivatives of $\fourg _{\mu\nu}$ on $\hyp$ in terms of  $(g,K)$ and their tangential derivatives. In this case the introduction of the index $k$ is clearly an overkill. However, we will see examples where each $k$ adds further information.

As a further example, consider two hypersurfaces $\mcNone$ and $\mcNtwo$ in $\mcM$ intersecting transversally at a joint boundary
$$
 \secN
  = \mcNone\cap \mcNtwo
 \,,
$$
and consider vacuum submanifold data $\CSdata{\mcNone\cup\mcNtwo,k}$ such that $\mcNone$ and $\mcNtwo$ are characteristic for $\fourg _{\mu\nu}$, with the null directions along $\mcNone$ and $\mcNtwo$ being orthogonal to $\secN$. Set
\begin{equation}\label{9I23.2}
  \Cdata{\mcN,k}=  \{ j^k_p g_{AB}, j^k_p \kappa, p\in \mcN\}
  \,,
\end{equation}
where $g_{AB}dx^Adx^B$
is the tensor field induced on $\mcN$ by $\fourg _{\mu\nu} dx^\mu dx^\nu$,
with $\kappa$ as in \eqref{4I23.11}  (transforming as in \eqref{15I23.7}), and with the data satisfying the Raychaudhuri equation and its transverse derivatives up to order $k$.
The collection \eqref{9I23.2} will be referred to as \emph{reduced characteristic data of order $k$} on $\mcN$.
It follows from Remark~\ref{R14I23.3} that, for $k\ge 1$,
vacuum characteristic data $\CSdata{\mcNone\cup\mcNtwo,k}$ can be replaced by a set
\begin{equation}\label{9I23.1}
  \Cdata{\mcNone ,0}\cup\Cdata{\mcNtwo ,0}\cup\CSdata{\secN ,1}
\end{equation}
together with the compatibility conditions and   constraints described in  Theorem~\ref{T5I23.1}. This last theorem also shows that every vacuum set $\CSdata{\mcNone\cup\mcNtwo,k}$ can be realised as the boundary of a manifold-with-boundary-with-corner carrying a smooth vacuum metric. Here the index $k$ is again an overkill for $k\ge 1$.

It is sensible to enquire about the minimum value of $k$ which makes sense in the context of Einstein equations. A first guess would be to take $k\ge 2$, since the equations are well posed in a classical sense when $k\ge 2$. However, some of the equations have a constraint character, involving two derivatives in tangential directions but only one  in  transverse directions. In view of our hypothesis, that all the fields are smooth in  tangential directions, an appropriate condition appears to be $k\ge 1$.

In the next sections we will:

\begin{enumerate}
  \item Provide a simpler description of vacuum characteristic   data $\Cdata{\mcN,k}$.
  \item Show that a class of vacuum characteristic   data $\Cdata{\mcN,k}$ can be realised by a hypersurface \emph{inside} a smooth vacuum spacetime.
      \item Show that vacuum spacelike data $\CSdata{\secN,k}$ on a submanifold $\secN$ of codimension two  can  be realised by a submanifold \emph{inside} a smooth vacuum spacetime
\end{enumerate}

\section{The Isenberg-Moncrief parameterisation}
 \label{s6I23.1}

\newcommand{\newdu}{\red{\mathrm{d} \newu}}
\newcommand{\newdr}{\red{\mathrm{d} \newr}}
\newcommand{\newu}{\red{u}}
\newcommand{\newr}{\red{r}}

A smooth submanifold $\secN$ of a smooth null hypersurface $\mcN\subset \mcM$ will be called a \emph{cross-section of $\mcN$} if every generator of $\mcN$ intersects $\secN$ precisely once.
The aim of this section is to provide a simpler description of characteristic data of order $k$ on a hypersurface $\mcN$ with cross-section $\secN$, using  coordinates which are often referred to as Gaussian null coordinates.
{
As  these coordinates have been introduced by Isenberg and Moncrief~\cite{VinceJimcompactCauchyCMP},
}
we will refer to them as  \emph{Isenberg-Moncrief coordinates}; IM for short. In these coordinates the metric reads%
\footnote{Our $r$ is the same as Isenberg-Moncrief's $t$ and our $u$ is the negative of Isenberg-Moncrief's $x^3$, see \cite[Equation~(2.8)]{VinceJimcompactCauchyCMP}.}
\begin{equation}
\label{10I22.w1}
 \fourg_{\mu \nu}\red{\mathrm{d}x}^\mu \red{\mathrm{d}x}^\nu
= 2 \left(- \newdu  + \newu \IMuu \newdr   + \newu \beta_{A} \red{\mathrm{d}x}^A \right) \newdr
         + \smet_{AB} \red{\mathrm{d}x}^A \red{\mathrm{d}x}^B \,.
\end{equation}

We let $\mcN$ be the hypersurface $\{u=0\}$, with $r$ being a coordinate along the generators of $\mcN$. The Einstein equations for the metric \eqref{10I22.w1}, in all dimensions, can be found in~\cite[Appendix A]{HIW}, after replacing the coordinates $(\newu,\newr) $  of~\cite{HIW} with our $(\newr,-\newu) $.

There are actually two choices for $\mcN$ in the coordinates \eqref{10I22.w1}: as $\{\newr=0\}$ or as $\{\newu=0\}$, since both these hypersurfaces are null for the metric \eqref{10I22.w1}.
The reader is warned that our  choice here leads unfortunately to confusions with the usual notation $\newu$ for a coordinate whose level sets are null, as is the case for Bondi coordinates: we emphasise that the zero-level set of $\newu$ here is null, but the remaining level sets are not, in general. On the other hand all the level sets of $r$ are null for a metric of the form \eqref{10I22.w1}. 

One can always  adjust the
Isenberg-Moncrief coordinates so that $\alpha$ vanishes on $\mcN$, but this
might not be convenient in general and will \emph{not} be assumed  in this section.

We shall  write $\beta^A := \smet^{AB}\beta_B$ and $\beta\cdot\beta:= \smet^{AB}\beta_A\beta_B$, where $\smet^{AB}$ is the inverse metric to $\smet_{AB}$.

In the coordinate system as in \eqref{10I22.w1} the key data on $\mcN$ are $(\smet_{AB},\IMuu)$ subject to the Raychaudhuri equation, which for the metric \eqref{10I22.w1} reads
\begin{equation}
    \label{10I22.w2}
    0 =
 \left(-\half \smet^{AB} \partial^2_\newr   \smet_{AB}
 +\quater \smet^{CA}\smet^{BD}(\partial_\newr  \smet_{AB})\partial_\newr  \smet_{CD}
 +\frac{1}{2} \IMuu \: \smet^{AB}\partial_\newr  \smet_{AB} \right)\bigg|_{\mcN}
\end{equation}
(see~\cite[Appendix~A]{HIW}, Equation (77) with their $\newr=0$). We will refer to the pair  $(\smet_{AB},\IMuu)$ as the \emph{Isenberg-Moncrief data on $\mcN$}. Note that  if $\theta\equiv \half \smet^{AB}\partial_\newr  \smet_{AB} $ has no zeros on $\mcN$, the field $\IMuu$ can be determined algebraically in terms of $\smet_{AB}$ and  derivatives of $\smet_{AB}$ tangential to $\mcN$  using \eqref{10I22.w2}.

{
\begin{remark}
 \label{R11III23.1}
{\rm  
It might of be interest to clarify how the Isenberg-Moncrief parameterisation \eqref{10I22.w1} of the metric fits with Theorem~\ref{T5I23.1}.
Either by a direct calculation, or by comparing   \eqref{21XII11.40} with \eqref{10I22.w2}  one finds
\begin{equation}\label{13I23.1ba}
  \alpha|_\mcN = \kappa_\mcN 
  \,.
\end{equation}
The constraint \eqref{21XII11.3} is trivially satisfied, 
while \eqref{21XII11.2} reads
\begin{equation}\label{13I23.1b}
  \alpha|_\secN = \kappa_\mcN 
  \,,
\end{equation}
consistently with \eqref{13I23.1ba}.
\qed
}
\end{remark}

In Isenberg-Moncrief coordinates the following holds:
}

\begin{enumerate}
  \item The equation $R_{\newr A}|_{\mcN}=0$ provides a
       linear ODE for $\beta_A$   along the generators of $\mcN$; as already mentioned,
       such ODEs will be referred to as \emph{transport equations along $\mcN$.}
       Indeed, in vacuum it holds
      that~\cite[Equation~(79)]{HIW}
\ptcheck{23I, for change of sign}
  \begin{eqnarray}
  \label{10I22.w3i}
  0 & = & R_{\newr A}
  \nonumber
\\
    &=&
  - D_A \IMuu
  + \half\partial_\newr \beta_A
  + \quater \beta_A \smet^{BC} \partial_\newr  \smet_{BC}
  - D_{[A}(\smet^{BC}\partial_\newr  \smet_{B]C})
\non \\
&& +
   \frac{\newu}{2}\times
   \bigg[\;
         \half (\smet^{BC}\partial_\newr \smet_{BC})\partial_\newu  \beta_A
         + \partial_\newr \partial_\newu \beta_A
         + 2\IMuu \partial_\newu \beta_A
\non \\
 && \qquad
         + \partial_\newu (\newu\beta_A) \cdot
                       \left\{
                             - \newu^{-1}\partial_\newu (\newu^{2}\beta\cdot\beta)
                              + 2 \partial_\newu \IMuu
                       \right\}
\non \\
 && \qquad
         - 2D_A(\partial_\newu \IMuu)
         + \partial_\newu (\beta^C\partial_\newr \smet_{CA})
         + 2\newu^{-1}\partial_\newu  \left(
                                     \newu^{2} \beta^D D_{[A}\beta_{D]}
                               \right)
\non \\
 && \qquad
         - \half \smet^{BC}\partial_\newu \smet_{BC}\times
           \Big\{
                   (\newu\beta\cdot\beta-2\IMuu)\partial_\newu (\newu\beta_A)
\non \\
&& \qquad \qquad \qquad \qquad \qquad
                  + 2 D_A \IMuu
                  - \beta^C\partial_\newr  \smet_{CA}
                  - 2 \newu \beta^F D_{[A}\beta_{F]}
           \Big\}
\non \\
 && \qquad
         - 2 \partial_\newu (\IMuu \beta_A)
         - 2 \newu(\partial_\newu  \IMuu) \partial_\newu  \beta_A
         - D_B\left\{\beta^B\partial_\newu (\newu\beta_A) \right\}
\non \\
 && \qquad
         + 2  \smet^{CD}D_D D_{[A}\beta_{C]}
         - \smet^{BC} (\partial_\newu \beta_B) \partial_\newr  \smet_{CA}
\non \\
 && \qquad
         - \smet^{BC}\partial_\newu (\newu\beta_B)\times
           \Big\{
                 - (\newu\beta\cdot\beta-2\IMuu) \partial_\newu \smet_{CA}
                  -  D_C \beta_A
\non \\
&& \qquad \qquad \qquad \qquad \qquad
                  - \beta_C\partial_\newu (\newu\beta_A)
                  + \newu\beta_C\beta^E\partial_\newu \smet_{EA}
           \Big\}
\non \\
 && \qquad
          + \smet^{BC}(\partial_\newu \smet_{CA})\times
             \left\{
                    2 \beta_B\partial_\newu (\newu\IMuu)
                    + 2 D_B \IMuu
                    + \partial_\newu  \beta_B
                    - 2 \newu \beta^D D_{[B}\beta_{D]}
             \right\}
   \bigg]
\,,
\phantom{xxx}
  \end{eqnarray}
  where $D_A$ is the covariant derivative operator associated with the metric $\smet_{AB}$.
  This equation at $\newu=0$,  together with $\beta_A|_{\secN}$ and the IM   data, determines $\beta_A$ uniquely along $\mcN$.
  Note that  only the first line of the right-hand side survives on $\mcN$, but we reproduce the relevant Einstein equations here and below in whole as the overall features of the remaining terms in the equations are relevant for the induction argument below.

Equations \eqref{22XII22.2}-\eqref{22XII22.5}, together with the equations obtained by differentiating \eqref{22XII22.2}-\eqref{22XII22.5} transversally to $\mcN$, will be referred to as \emph{the transport equations along $\mcN$.}

  \item The vacuum equation $R_{AB}|_{\mcN}= - \frac{n+3}{2(n-1)}\Lambda \smet_{AB}$
  provides a linear transport equation for $\partial_\newu  \smet_{AB}$.
  Indeed, we have~\cite[Equation~(82)]{HIW}
 \ptcheck{13I23, for change of sign}
\begin{eqnarray}
\label{10I22.w4}
R_{AB}
        &=&
          \partial_\newr \partial_\newu  \smet_{AB}
         + \IMuu \partial_\newu  \smet_{AB}
         +  {\cal R}_{AB}
         - D_{(A}\beta_{B)}
         - \half \beta_A\beta_B
\non \\
        &-&
           \smet^{CD} \left(\partial_\newu \smet_{D(A}\right)\partial_\newr  \smet_{B)C}
         + \quater
           \left\{
                   (\smet^{CD}\partial_\newr  \smet_{CD})\partial_\newu  \smet_{AB}
                 +(\smet^{CD}\partial_\newu  \smet_{CD})\partial_\newr  \smet_{AB}
           \right\}
\non \\
 &-& \frac{\newu}{2} \times
     \bigg[
           - 2\IMuu \partial_\newu ^2 \smet_{AB}
           + D_C(\beta^C\partial_\newu  \smet_{AB})
\non \\
  && \qquad \,
           + \half (\smet^{CD}\partial_\newu \smet_{CD})
             \left\{
                     (\newu\beta\cdot\beta-2\IMuu)\partial_\newu \smet_{AB}
                   + 2 D_{(A}\beta_{B)}
             \right\}
\non \\
  && \qquad \,
           - 2(\partial_\newu  \IMuu) \partial_\newu  \smet_{AB}
           + \newu^{-1}\{\partial_\newu (\newu^{2}\beta\cdot\beta)\}
                   \partial_\newu  \smet_{AB}
\non \\
  && \qquad \,
           + \newu\beta\cdot\beta \partial_\newu ^2 \smet_{AB}
           + 2\partial_\newu  \{ D_{(A}\beta_{B)}  \}
\non \\
  && \qquad \,
           + 2\beta_{(A}\partial_\newu \beta_{B)}
           + \newu (\partial_\newu \beta_A) \partial_\newu \beta_B
           + \newu\beta^E\beta^F
             (\partial_\newu  \smet_{AE})\partial_\newu \smet_{BF}
\non \\
  && \qquad \,
           - 2\beta^C
             \left\{\partial_\newu (\newu\beta_{(A}) \right\}\partial_\newu \smet_{B)C}
           - 2\smet^{CD}\left(D_D \beta_{(A} \right)
             \partial_\newu \smet_{B)C}
\non \\
  && \qquad \,
           - \smet^{CD} (\newu \beta\cdot\beta-2\IMuu)
                      (\partial_\newu \smet_{CA})\partial_\newu \smet_{BD} \,
     \bigg]
     \,,
\end{eqnarray}
where ${\cal R}_{AB}$ is the Ricci tensor associated with $\smet_{AB}$.
This equation at $\newu=0$, together with $\partial_\newu  \smet_{AB} |_{\secN}$ and the field $\beta_A|_\mcN$ determined in point 1., defines now uniquely  $\partial_\newu  \smet_{AB}|_{\mcN}$.
\item The equation $R_{\newu\newr}|_{\mcN}= -\frac{n+3}{2(n-1)}\Lambda$, where~\cite[Equation~(78)]{HIW}
\ptcheck{22I, by wan}
\begin{eqnarray}
\label{10I22.w5}
 R_{\newu\newr}
     &=& -2 \partial_\newu \IMuu
         + \quater \smet^{CA}\smet^{BD}(\partial_\newr \smet_{CD})\partial_\newu \smet_{AB}
         -\half \smet^{AB} \partial_\newu  \partial_\newr  \smet_{AB}
         -\frac{1}{2} \IMuu\: \smet^{AB} \partial_\newu \smet_{AB}
         +\half \beta^A\beta_A
\non \\
 &+&
   \frac{\newu}{2}\times
    \bigg[
          -2\partial_\newu ^2\IMuu
          -\half \smet^{AB}\partial_\newu \smet_{AB} \cdot
           \left\{
                  2\partial_\newu \IMuu - \beta^D\partial_\newu (\newu\beta_D)
           \right\}
\non \\
 && \qquad \, \,
          + \beta^B\partial_\newu \beta_B
          + \partial_\newu \{ \beta^B\partial_\newu (\newu\beta_B)\}
          + \smet^{AB}D_A(\partial_\newu \beta_B)
    \bigg] \,,
\end{eqnarray}
%
determines now algebraically $\partial_\newu \IMuu|_{\mcN}$, as well as any derivative  of $\partial_\newu  \IMuu|_{\mcN}$ in directions tangential to $\mcN$.
\item The transverse derivative  $ \partial_\newu   \beta_A|_{\mcN} $, as well as  the derivatives of $ \partial_\newu   \beta_A|_{\mcN} $  tangential  to $\mcN$, are now obtained from the equation $R_{\newu A}|_{\mcN}=0$, where~\cite[Equation~(81)]{HIW}
\begin{eqnarray}
    \label{10I23.w6}
    R_{\newu A}
      &=&
         - \quater \beta_A \smet^{BC}\partial_\newu  \smet_{BC}
         - \partial_\newu  \beta_A
         + \half \beta^B\partial_\newu  \smet_{AB}
         - D_{[A}
           \left(\smet^{BC}\partial_\newu  \smet_{B]C} \right)
\non \\
 & + &  \frac{\newu}{2}\times
        \bigg[\;
                - \partial_\newu ^2 \beta_A
               + \partial_\newu
                \left(\beta^B\partial_\newu \smet_{AB} \right)
\non \\
&& \qquad \qquad
                + \half (\smet^{CD}\partial_\newu  \smet_{CD})
                    \left(
                          - \partial_\newu \beta_A
                          + \beta^B\partial_\newu  \smet_{AB}
                    \right)
        \bigg] \,.
\end{eqnarray}
%
\end{enumerate}

One can now consider the equations obtained by successively applying $\partial_\newu ^i$, $i\ge 1$, to
\eqref{10I22.w4}-\eqref{10I23.w6}
to determine all transverse derivatives of $\smet_{AB}$ on $\mcN$
   by first order linear transport equations, whose solutions are determined uniquely by their initial or final data,
and all transverse derivatives of  $\IMuu$ and $\beta_{A}$ on $\mcN$ from linear algebraic equations.

We have therefore proved:

\begin{Proposition}
  \label{P11I23.1}
In the Isenberg-Moncrief coordinate system the reduced
vacuum characteristic data
$\Cdata{\mcN,k}$ of \eqref{9I23.2} can be replaced by
the following collection of fields:
\begin{equation}\label{8I23.5a}
 \CdataIM{\mcN,k}:= \{  (\partial_\newu ^j \smet_{AB},  \IMuu, \beta_A)_{0\le  j \le k} \ \mbox{on} \ \secN \ \mbox{and}
 \
  (\smet_{AB},\IMuu) \ \mbox{on} \ \mcN
  \}
  \,,
\end{equation}
with $(\smet_{AB},\IMuu)$ subject to the  constraint  equation \eqref{10I22.w2}.   Moreover the fields $ (\alpha,\beta_A)$ are  not needed in \eqref{8I23.5a} if $k=0$, as they appear multiplied by $\newu$ in the metric.
\qed
\end{Proposition}

We will refer to this set of data as \emph{Isenberg-Moncrief
vacuum characteristic data
 of order $k$.}

The above also shows: 

\begin{Proposition}
  \label{P11I23.2}
  Let $\secN$ be of codimension-two and $k\in \N\cup\{\infty\}$. Then:

 \begin{enumerate}
   \item
   Spacelike vacuum  data  $\CSdata{\secN,k}$ of \eqref{8I23.7a} can be reduced to
the following collection of fields on $\secN$ in Isenberg-Moncrief coordinates:
 \ptcheck{I23, together, and confirmed again, that \eqref{21XII11.2} does not give new constraint.
}
\begin{equation}\label{8I23.5b}
 \CSdataIM{\secN,k}:=  (\partial_r^i\partial_u ^j \smet_{AB},  \IMuu, \beta_A)_{0\le i+j \le k}
  \,,
\end{equation}
 with $(\alpha,\beta_A)$ not needed if $k=0$.
   \item
Let $\secN_1$ and $\secN_2$ be two cross-sections of a null hypersurface $\mcN$ with vacuum  data  $\CSdata{\secN_1,k}$ and $\CSdata{\secN_2,k}$ induced by vacuum characteristic data $\CSdata{\mcN,k}$. Then $\CSdata{\secN_2,k}$ can be determined uniquely in terms of $\CSdata{\secN_1,k}$ and $\CSdata{\mcN,0}$
 by solving linear transport equations along the generators of $\mcN$, or by solving linear algebraic equations on $\mcN$; and vice-versa.
 \end{enumerate}
\end{Proposition}

\proof
1. This can be seen directly by repeating the proof of Proposition~\ref{P11I23.1}. Alternatively, let $\fourg _{\mu\nu}$ be any Lorentzian metric near $\secN$ compatible with $\CSdata{\secN,k}$.
Denote by $\mcNone$ either  of the null hypersurfaces orthogonal to $\secN$, and by $\mcNtwo$ the other one. Use the data induced by $\fourg _{\mu\nu}$ to solve the vacuum Einstein equations to the future of $\mcNone\cup\mcNtwo$. Introduce Isenberg-Moncrief coordinates so that $\mcNone=\{\newu=0\}$. The result follows now from Proposition~\ref{P11I23.1}.

2. It suffices to prove the result in Isenberg-Moncrief coordinates. In these we have just shown that one can reduce  $\CSdata{\secN_1,k}$ to $\CSdataIM{\secN_1,k}$  and $\CSdata{\secN_2,k}$ to $\CSdataIM{\secN_2,k}$, with $\CSdataIM{\secN_2,k}$  determined uniquely in terms of $\CSdataIM{\secN_2,k}$ and $\CSdata{\mcN,0}$ by solving linear transport equations along the generators of $\mcN$ and by solving linear algebraic equations on $\mcN$, as desired.
\qed

\section{The Bondi parameterisation}
 \label{s22XII22.2}

A  parameterisation of the metric which has often been used in the literature,
 in spacetime-dimension equal to four,
 is that of
Bondi et al.\ (cf., e.g., \cite{Sachs,BBM,Frittelli:2004pk,Frittelli:1999yr,MaedlerWinicour}),
\begin{eqnarray}
 \fourg &\equiv& \fourg_{\alpha\beta}\mathrm{d}x^{\alpha}\mathrm{d}x^{\beta}
 \nonumber
\\
 &  =  &-\frac{V}{r}e^{2\beta} \mathrm{d}u^2-2 e^{2\beta}\mathrm{d}u \mathrm{d}r
   +r^2\zhTBW_{AB}\Big(\mathrm{d}x^A-U^A\mathrm{d}u\Big)\Big(\mathrm{d}x^B-U^B\mathrm{d}u\Big)
    \, ,
     \label{22XII22.1}
\end{eqnarray}
together with the conditions
\begin{equation}\label{26XII22.1a}
  \partial_r \det{\zhTBW_{AB}}=
  \partial_u\det{\zhTBW_{AB}}=
   0
    \,.
\end{equation}
A coordinate system satisfying \eqref{26XII22.1a} exists on $\mcN$ with $\mcN = \{u=0\}$ if and only if the expansion scalar of $\mcN$ has no zeros.  We show in Appendix~\ref{s18IV22.1asdf},
 in all spacetime dimensions $n+1\ge 3$,
  that given a  cross-section $\secN$ of a smooth null hypersurface $\mcN$ with $\thetatau>0$,
there exists a unique Bondi coordinate system
near $\mcN$ in which
\begin{equation}\label{4I23.22}
\mbox{
 $u|_{\mcN} =0$, $r|_{\secN}=1$, $\thetatau|_{\secN}=n-1$, $V|_{\secN} = - \frac{2\uthetatau}{n-1}$  and
  $\beta|_{\secN}=0= \mcU^A|_{\secN}$.
  }
\end{equation}
Here $\thetatau$ is the null mean curvature of $\secN$ associated with  $\partial_r$,
and $\uthetatau$ is the null mean curvature of   $\secN$ associated with the null vectors, say $\underline{\check L}$,
 orthogonal to $\secN$ and transverse to $\{u=0\}$ at $\secN$, normalised so that
$\fourg (\underline{\check L},\partial_r)|_{\secN}  = -1$.

\medskip

{
\begin{Remark}\label{R14I23.1}
{\rm
Some comments concerning Theorem~\ref{T5I23.1} and the Bondi form of the metric \eqref{22XII22.1}
are in order. Suppose  that  $\thetatau_\mcNone\ne 0$ in  Theorem~\ref{T5I23.1}.
We can then introduce on $\mcNone$ an area coordinate $r$ (compare Appendix~\ref{s18IV22.1asdf}), and use the Bondi parameterisation of the metric on $\mcNone$. So we take $\xtwo|_\mcNone = r$, and we emphasise that  \eqref{29X0.1a1} will \emph{not} hold in general when $\xtwo$ there is taken to be the Bondi area coordinate. Thus the Bondi form of the metric is assumed to hold on $\mcNone$ but not necessarily away from $\mcNone$.%
\footnote{One could be tempted to think that those of the Bondi-parameterised Einstein equations which \emph{do not} contain $\partial_u$-derivatives of the metric can be used \emph{as they are} on $\mcNone$, but this is not the case: for instance, the Raychaudhuri equation (compare~\eqref{7I23.1})   contains a term which involves $\partial_u$-derivatives of the metric and which vanishes if the Bondi form of the metric holds in a neighborhood of $\mcN$, and which does not vanish in a general coordinate system.
}

Using the Bondi parameterisation on $\mcNone$ we have
\begin{eqnarray}\label{6I23.1}
 & \displaystyle
 \xtwo|_\mcNone = r \,,
 \qquad
  \thetatau_\mcNone = \frac{n-1}{r}
  \,,
  \qquad
  \sigma_{AB}  = \frac{r^{2}}{2}\partial_r \gamma_{AB}|_\mcNone
   \,.
    &
\end{eqnarray}
The constraint equation \eqref{21XII11.40} can be solved for $\kappa_{\mcNone}$:
\begin{equation}
         \label{7I23.11}
 \kappa_{\mcNone} = \frac{r}{4(n-1)}\zhTBW^{AC}\zhTBW^{BD} (\partial_{r} \zhTBW_{AB})(\partial_r \zhTBW_{CD})\big|_\mcNone
    \,.
\end{equation}
Equations \eqref{21XII11.2} and \eqref{7I23.11} provide then a constraint on $\secN$:
\begin{eqnarray}
 & \displaystyle
 \Big[ 2 \partial_r\beta
   + \frac{e^{-2\beta}}{2}
    \overline{ \partial_\xone \red{g}_{\xtwo \, \xtwo}  }
    \Big]\Big |_\secN
    =\kappa_\mcNone |_\secN = \frac{r}{4(n-1)}\zhTBW^{AC}\zhTBW^{BD} (\partial_{r} \zhTBW_{AB})(\partial_r \zhTBW_{CD})\big|_\secN
  \,.
  &
 \label{8I23.1}
\end{eqnarray}

Once the vacuum equations have been solved, $\kappa_\mcNone$ coincides with $\Gamma^\xtwo _{\xtwo\xtwo}$
restricted to $\mcNone$:
\begin{eqnarray}
 & \displaystyle
    \kappa_\mcNone
    =  2 \partial_r\beta
   + \frac{e^{-2\beta}}{2}
    \overline{ \partial_\xone \red{g}_{\xtwo \, \xtwo}  }
  \,.
  &
 \label{8I23.4}
\end{eqnarray}
Hence the vacuum Raychaudhuri equation \eqref{21XII11.40}
becomes
\begin{equation}
         \label{7I23.1}
  2 \partial_{r} \beta
  \underbrace{+  \frac{e^{-2\beta}}{2}\overline{ \partial_\xone \red{g}_{\xtwo \, \xtwo}}
   }_{\phantom{x}} = \frac{r}{4(n-1)}\zhTBW^{AC}\zhTBW^{BD} (\partial_{r} \zhTBW_{AB})(\partial_r \zhTBW_{CD})  \Big |_\mcNone
    \,.
\end{equation}
This is consistent  with \eqref{22XII22.2} below, since the underbraced term is zero when $\xtwo$ is taken to be the area coordinate $r$ and the Bondi form of the metric is assumed in a neighborhood of $\mcNone$.
\qed

}
\end{Remark}
}

In the notation of \eqref{22XII22.1},  cross-section data can be defined as follows. Let $\secN$ be a cross-section of $\mcN$. Let $k \in \N\cup\{\infty\}$ be the number of \emph{transverse} derivatives of the metric that we want to control at $\secN$.  Using the  Bondi parameterisation of the metric, we define the \emph{Bondi cross-section data of order $k$} as the collection of fields
\begin{equation}\label{23III22.99}
  \CSdataBo{\secN,k}:=(\partial_r^i\partial_u^{j}\gamma_{AB}|_{\secN}
  ,\,  \partial_u^{j}\beta|_{\secN},
  \, \partial_u^{j}U^A|_{\secN}
  ,\, \partial_r U^A|_{\secN}
  ,\,    V|_{\secN}
  )_{0\le i+j\leq k}
  \,.
\end{equation}
As already pointed out, for simplicity we assume that all the  fields in \eqref{23III22.99}
are smooth.
We note that a finite sufficiently large degree of differentiability, typically different for distinct fields, would suffice for most purposes; this can be determined by chasing the number of derivatives needed in the relevant equations.

In Appendix~\ref{app22XII22.1} we show how the sphere data of \cite{ACR1} relate to the Bondi ones.

We will refer to the Bondi field $\gamma_{AB}|_\mcN$ as  \emph{free Bondi data} on $ \mcN$.

Note that the data $\CSdataBo{\secN,k}$  contain more fields than the Isenberg-Moncrief data
 $\CSdataIM{\secN,k}$ of \eqref{8I23.5b}. This is related to the residual freedom remaining in  Bondi coordinates; compare Proposition~\ref{R25I23}, Appendix~\ref{s18IV22.1asdf}.

To continue, suppose that we are given Bondi cross-section data of order $k$ and a field $\gamma_{AB}$ on $\mcN$. We assume that the field $\gamma_{AB}$  and its $\partial^j_r$-derivatives with $0\le j\le k$, when restricted to $\secN$, coincide with the Bondi cross-section fields  $\partial^j_r\gamma_{AB}|_{\secN}$; such data sets will be said to be compatible. Then (see~\cite{MaedlerWinicour} in spacetime-dimension four):

\begin{enumerate}
 \item Using the equation
\begin{equation}
         \label{22XII22.2}
0 = \frac{r}{2(n-1)} G_{rr} = \partial_{r} \beta - \frac{r}{8(n-1)}\zhTBW^{AC}\zhTBW^{BD} (\partial_{r} \zhTBW_{AB})(\partial_r \zhTBW_{CD})
\end{equation}
we can determine  $\partial_r \beta|_{\secN}$.
Further, integrating \eqref{22XII22.2}, the value  $ \beta|_{\secN}$ and the field $\gamma_{AB}$ on $\mcN$ can be used to determine $\partial_r\beta$ on $\mcN$, and hence all radial derivatives $\partial_r^i\beta|_{\secN}$.

 \item    The fields $U^A|_{\secN}$ and $\partial_rU^A|_{\secN}$  are used to obtain $U^A|_{\mcN}$ by integrating
\begin{eqnarray}
        0
        &= &
            2r^{n-1}  G_{rA}
             \nonumber
\\
             &= &
              \partial_r \left[r^{n+1} e^{-2\beta}\zhTBW_{AB}(\partial_r U^B)\right]
            -
            2r^{2(n-1)}\partial_r \Big(\frac{1}{r^{n-1}}\spaceD_A\beta  \Big)
                 +r^{n-1}\zhTBW^{EF} \spaceD_E (\partial_r \zhTBW_{AF})
                 \,.
                 \nonumber
                 \\
                            \label{22XII22.3}
           \end{eqnarray}
Further,  the radial derivatives $\partial_r^iU^A|_\secN$ can be algebraically determined  in terms of $\partial_r^i\gamma_{AB}|_\secN$.

We note that $U_A$ can be set to zero on $\secN$ by a refinement  of the coordinates, but this restriction is not convenient when  interpolating between cross-section data, and will therefore not be assumed.
 \item
  The function $V|_{\secN}$  is needed to integrate the equation
 \begin{eqnarray}
         2 \Lambda  r^2
                   &=&
               r^2 e^{-2\beta} (2 G_{ur} + 2 U^A G_{rA} - V/r\, G_{rr} )
               \nonumber
\\
               & = &
                 R[\zhTBW]
                -2\zhTBW^{AB}  \Big[\spaceD_A \spaceD_B \beta
                + (\spaceD_A\beta) (\spaceD_B \beta)\Big]
                +\frac{e^{-2\beta}}{r^{2(n-2)} }\spaceD_A \Big[ \partial_r (r^{2(n-1)}U^A)\Big]
               \nonumber
\\
                &&
                 -\frac{1}{2}r^4 e^{-4\beta}\zhTBW_{AB}(\partial_r U^A)(\partial_r U^B)
                -\frac{(n-1)}{r^{n-3}} e^{-2\beta} \partial_r( r^{n-3} V)
                                 \,,
                  \label{22XII22.4}
           \end{eqnarray}
obtaining thus $V|_{\mcN}$ by integration.
Further, all radial derivatives $\partial^i _r V|_{\secN}$ can be determined algebraically in terms of the already-known fields by inductively $r$-differentiating this equation.

 \item     The field $\partial_u \gamma_{AB}|_{\secN}$ is  needed to integrate
\begin{eqnarray}
 0
 & = &   r^{(n-5)/2} \TS[ G_{AB}]
 \nonumber
\\
& = &
    \partial_r
    \Big[
     r^{(n-1)/2}   \partial_u \zhTBW_{AB}
     	 - \frac{1}{2} r^{(n-3)/2 } V   \partial_r \zhTBW_{AB}
     	 -  \frac{n-1}{4} r^{(n-5)/2  } V    \zhTBW_{AB}
     \Big]
           \nonumber
\\
 && +  \frac{n-1}{4} \partial_r(r^{(n-5)/2  } V )   \zhTBW_{AB}
  \nonumber
  \\
       &&
      \fbox{$ +
       \displaystyle
       \frac{1}{2} r^{(n-3)/2}
        V \gamma^{CD} \partial_r\gamma_{AC}\partial_r\gamma_{BD}
       -\frac{1}{2}
        r^{(n-1)/2}
        \gamma^{CD}(\partial_r\gamma_{BD}\partial_u\gamma_{AC}
       +\partial_u\gamma_{BD}\partial_r\gamma_{AC})
       $
       }
       \nonumber \\
       &&
        +
       r^{(n-5)/2}\TS
       \bigg[
      e^{2\beta} r^2  R[\gamma]_{AB}  -2e^{\beta} \spaceD_A \spaceD_B e^\beta
      \nonumber
\\
 &&
      + r^{3-n} \zhTBW_{CA} \spaceD_B[ \partial_r (r^{n-1}U^C) ]
          - \frac{1}{2} r^4 e^{-2\beta}\zhTBW_{AC}\zhTBW_{BD} (\partial_r U^C) (\partial_r U^D)
          \nonumber \\
       &&
       +
               \frac{r^2}{2}  (\partial_r \zhTBW_{AB}) (\spaceD_C U^C )
              +r^2 U^C \spaceD_C (\partial_r \zhTBW_{AB})
                \nonumber \\
       &&
        -
	r^2 (\partial_r \zhTBW_{AC}) \zhTBW_{BE} (\spaceD^C U^E -\spaceD^E U^C)
       \bigg]
       \,,
       \label{22XII22.5}
\end{eqnarray}
where the symbol $\TS$ denotes the traceless-symmetric part of a tensor with respect to the metric $\gamma_{AB}$ and where $R[\gamma]_{AB}$ is the Ricci tensor of the metric $\gamma_{AB}$; note that the TS part of each of the terms in the box  is zero when $n=3$. One thus determines
 $\partial_u \gamma_{AB}|_{\mcN}$.
\item
The   $u$-derivative of $\beta$ on $\mcN$ can  be calculated by integrating the equation obtained by $u$-differentiating  \eqref{22XII22.2}, after expressing the right-hand side  in terms of the fields determined so far:
\begin{equation}
         \label{22XII22.11}
          \partial_r \partial_u \beta  = \frac{r}{8 }
          \Big(
           \zhTBW^{AC}\partial_u\zhTBW^{BD} (\partial_r \zhTBW_{AB})(\partial_r \zhTBW_{CD})
           +
           \zhTBW^{AC}\zhTBW^{BD} (\partial_r \zhTBW_{AB})(\partial_r\partial_u \zhTBW_{CD})
           \Big)
          \,.
\end{equation}
For this we also need the initial value   $\partial_u  \beta|_{\secN} $.

\item
 \ptcheck{6I23 together with mathematica file}
The equation
\begin{eqnarray}
        -2 e^{2\beta}G_{uA}  =0
        \,,
\end{eqnarray}
reads
\begin{eqnarray}
\lefteqn{
   \partial_r \Big[
        e^{4\beta}\partial_u\Big(e^{-4\beta}r^2\gamma_{AB}U^B\Big)
    \Big]
        +\partial_r(r^2 \gamma_{AB})\partial_u U^B
      }
 & &
      \nonumber
\\
       & = &
        \partial_r \Big[
         e^{2\beta} \partial_r \Big( r\gamma_{AB}U^B V e^{2\beta}\Big) -2 r V \partial_r (\gamma_{AB}U^B)
    + r^2 U^B\partial_u\gamma_{AB}
    \Big]
   +
    2 e^{2\beta} \partial_A\partial_u\beta
 \nonumber
\\
     &&
     + \Big(U^B \partial_r\big(r^2\gamma_{AB}\big) + 1/2 \gamma_{AB}r^2 \partial_r U^B\Big)
      e^{4\beta}\partial_u  e^{4\beta}
    -   2  \Lambda r^2  e^{2\beta}\gamma_{AB} U^B +
 F_A
     \,,
      \phantom{xxxxxx}
      \label{1V22.12}
\end{eqnarray}
with
\begin{equation}\label{1V22.13}
  F_A = F_A\big( {n},r,\fourg_{\mu\nu}, \partial_i\fourg _{\mu\nu}, \partial_u \gamma_{AB}, \partial_A \partial_B\fourg _{\mu\nu}
  , \partial_r \partial_B\fourg _{\mu\nu} , \partial_A\partial_u\gamma_{AB}
  \big)
  \, ,
\end{equation}
where $\{\partial_i\}= \{\partial_r,\partial_A\}$. This equation
allows us to determine algebraically $\partial_r\partial_u U^A|_\secN$ in terms of the already known fields.
 The field $\partial_r\partial_u U^A|_\secN$ is needed to determine $\partial_u U^A|_\mcN$  by integrating in $r$ the $u$-derivative of Equation~\eqref{22XII22.3}.
The explicit form of $F_A$ is not very enlightening and is too long to be usefully displayed here.

 \item We can determine algebraically  $\partial_uV$ either on $\secN$ or on $\mcN$  from the Einstein equation $
     (G_{uu}+\Lambda g_{uu})|_\mcN = 0$:
  \begin{equation}\label{13I23.1}
    G_{uu} = \frac{n-1}{2 r^{2}}\partial_u V + ...
    \,,
  \end{equation}
  where ``$...$'' stands for an explicit expression in all fields already known on $\mcN$, and which is too long to be usefully displayed here.
\end{enumerate}
One can  inductively repeat the procedure above using the equations obtained by differentiating Einstein equations with respect to $u$. One thus obtains a hierarchical system of  ODEs in $r$, or algebraic equations for the transverse derivatives of
$\partial_r U^A$ and $V$,
for
\begin{equation}\label{27XII22.3}
(\beta ,\, U^A ,\,
 V ,\,  \partial_u  \gamma_{AB} ,\,  \partial_u \beta
  ,\,\partial_u U^A ,\,  \partial_u V,\,  \partial_u^2  \gamma_{AB} ,
  \ldots
  ,\,
\partial_u^k \beta ,\,\partial_u^k U^A ,\,  \partial^k_u V)
\,,
\end{equation}
which can be integrated or algebraically solved in the order indicated in \eqref{27XII22.3}.

It might be of some interest to note that there are no obstructions to integrate the transport equations globally on $I\times \secN$. Here one should keep in mind that \eqref{22XII22.2} is the Raychaudhuri equation written in terms of the Bondi fields, and the blow-up of solutions of this equation is at the origin  of various incompleteness theorems in general relativity.

Summarising, we have:

\begin{Proposition}
  \label{P11I23.1Bo}
In the Bondi coordinate system (the existence of which requires $\theta>0$ on $\mcN$) the reduced
vacuum characteristic data
$\Cdata{\mcN,k}$ of \eqref{9I23.2} can be replaced by
the following collection of fields:
\begin{equation}\label{8I23.5aBo}
 \CdataBo{\mcN,k}:= \{  ( \partial_u^{j}\gamma_{AB}|_{\secN}
  ,\,  \partial_u^{j}\beta|_{\secN},
  \, \partial_u^{j}U^A|_{\secN}
  ,\, \partial_r U^A|_{\secN}
  ,\,    V|_{\secN}
  )_{0\le j\leq k} \ \mbox{on} \ \secN \ \mbox{and}
 \
  \gamma_{AB}  \ \mbox{on} \ \mcN
  \}
  \,.
\end{equation}
Equivalently,
Einstein equations and smooth cross-section Bondi data \eqref{23III22.99} of order $ k\in\N\cup\{\infty\}$,
 together with smooth compatible Bondi free data $\gamma_{AB}$ on a null hypersurface $\mcN$,  determine uniquely the metric functions
$$
  \partial_u^i\fourg _{\mu\nu} |_{\mcN}
$$
in  Bondi gauge, for all  $0\le i \le k$,
 through linear transport equations along the generators, or through linear algebraic equations.

 Furthermore,  in vacuum, the Bondi data $\CSdataBo{\secN,k}$ of \eqref{23III22.99} suffice  to determine  $\CSdata{\secN,k}$ in Bondi coordinates.
 \qed
\end{Proposition}

{\redc
}

\section{The ``hand-crank construction''}
 \label{s26XII22.1}

As already pointed out, characteristic initial data on a single null hypersurface do not lead to a well posed Cauchy problem, i.e., a setup which guarantees both existence and uniqueness of associated solutions of the field equations. In this section we present a construction which provides existence of solutions of the vacuum Einstein equations realising the data.
A similar idea can be found in  \cite{Kehle:2022uvc}.

We have:

\begin{Proposition}
 \label{P26XII22.1}
Given a smooth vacuum characteristic initial data set $\noCdata{\mcN,k}$,  $k\in \N\cup\{\infty\}$, on a
{
$(n+1)$-dimensional manifold, $n\ge 3$,
}
of the form
$$
 \mcN=  [r_1,r_2]\times \secN\,,\qquad r_1< r_2
 \,,
$$
there exist (many) smooth solutions of the vacuum Einstein equations $(\mcM,\fourg)$ so that $\noCdata{\mcN,k}$ is  obtained by restriction to a characteristic hypersurface within $\mcM$.
\end{Proposition}

\proof
If $k<\infty$ we can extend $\noCdata{\mcN,k}$ to  $\noCdata{\mcN,\infty}$ in any way. Therefore it suffices to assume that $k=\infty$.

\begin{figure}
  \centering
\includegraphics[scale=0.38]{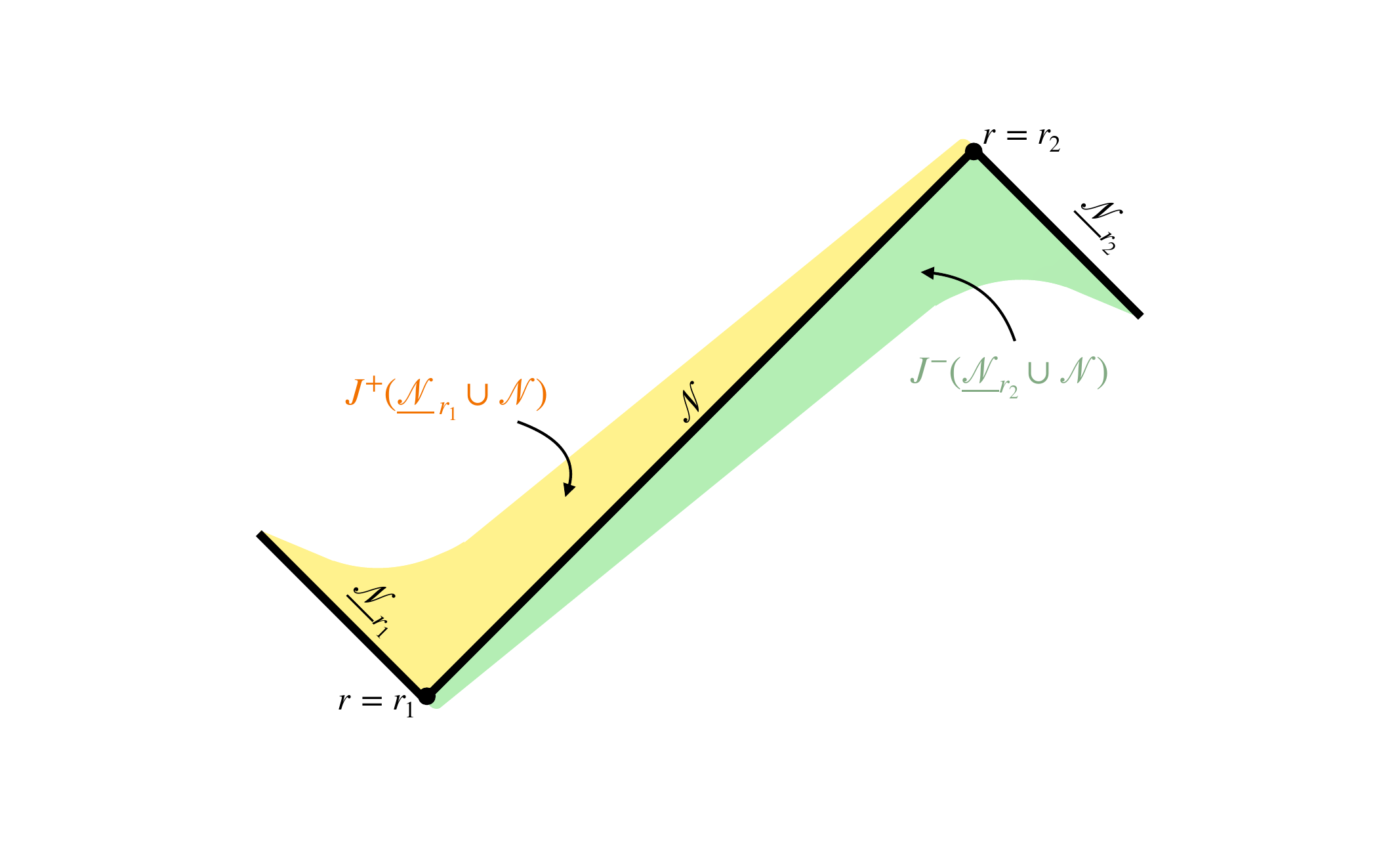}
\caption{The ``hand-crank construction''.
}
   \label{F24XII22.1}
\end{figure}

Let us write $\secN_c$ for $\mcN\cap \{r=c\}$. Let $\CSdata{\secN_{r},k}$ be the cross-section data induced by  $\noCdata{\mcN,k}$ on $\secN_r$. Let $\whmcN{r_1}=  [0,1]\times \secN$ be a hypersurface on which we prescribe any smooth Isenberg-Moncrief data   $( \smet_{AB}, \alpha)$
compatible with $\CSdata{\secN_{r_1},k}$;
thus $(\whmcN{r_1}, \smet_{AB},  \alpha)$ are characteristic initial data on the  hypersurface
$\whmcN{r_1}$ meeting $\mcN$ transversally at $r=r_1$ towards the \emph{future} of $\secN_{r_1}$, see Figure~\ref{F24XII22.1}. By definition of compatibility, all  derivatives  of $(\smet_{\mu\nu},\alpha)$ in directions tangential to $\whmcN{r_1}$ and transverse to $\mcN$ have to match at $\secN_{r_1}$ with those in $\CSdata{\secN_{r_1},k}$,
which can be achieved by Borel summation.

Similarly
let $\whmcN{r_2}=  [0,1]\times \secN$ be a hypersurface, meeting $\mcN$ transversally at $\secN_{r_2}$ towards the \emph{past}, on which we give smooth characteristic data compatible with $\CSdata{\secN_{r_2},k}$.

We can solve the characteristic Cauchy problem to the future with data on the transversally intersecting hypersurfaces
$ \mcN $
and $\whmcN{r_1}$, resulting in a smooth vacuum metric, say $\fourg_+$, defined on
$J^+(\whmcN{r_1}\cup\mcN)$. The Einstein equations guarantee that the transport  and the algebraic equations  in Isenberg-Moncrief coordinates
described in   Section~\ref{s6I23.1} hold on $\mcN$. The initial data for these transport equations at $\secN_{r_1}$ are given by $\CSdata{\secN_{r_1},k}$. Uniqueness of solutions of the transport equations shows that the transverse derivatives of the metric coincide with the ones listed in $\noCdata{\mcN,k}$.

One can  likewise solve the characteristic Cauchy problem to the past with data on the transversally intersecting hypersurfaces $\mcN $ and $\whmcN{r_2}$, resulting in a vacuum metric, say $\fourg_-$, defined on
$J^-(\mcN \cup \whmcN{r_2})$. Uniqueness of solutions of the transport equations shows that the transverse derivatives of the metric coincide with the ones listed in $\noCdata{\mcN,k}$.

By construction both piecewise-smooth spacetime metrics agree, together with all transverse derivatives, on $\mcN{}$, and define thus a smooth metric   on the union of the original domains of definition.
 \qedskip

As a corollary of the construction of Proposition~\ref{P26XII22.1} we have:

\begin{Corollary}
 \label{C26XII22.1}
 Let $k\in\N\cup\{\infty\}$, $k\ge 2$, and let $\secN\subset \mcM$ be of codimension two.
 For every spacelike vacuum data set $\CSdata{\secN,k}$ there exists  a smooth vacuum Lorentzian metric defined near $\secN$ and inducing the data.
\end{Corollary}

\proof
Let $\mcN=[-1,1]\times \secN$, where we identify $\secN$ with $\{0\}\times \secN$. Let $\noCdata{\mcN,k}$ be any characteristic data on $\mcN$ compatible with $\CSdata{\secN,k}$ at  $\{0\}\times \secN$.
If $k<\infty$, we complement  the data with higher-derivatives to $k=\infty$ in any way. Hence it suffices again to assume that $k=\infty$. We can now apply Proposition~\ref{P26XII22.1} to $\noCdata{\mcN,k}$.
\qed

{
\begin{Remark}
 \label{R28V23.1}
\rm
We note that in situations where Cauchy-stability holds but only a neighborhood of $\mcNone\cap\mcNtwo$ (instead of $\mcNone\cup\mcNtwo$) is known to exist (which could be the case for Einstein equations coupled with some unusual matter fields), one can still establish the claim of Corollary~\ref{C26XII22.1} as follows:
\color{black}
There exists $k_0>0$ so that Cauchy-stability for the characteristic initial value problem holds  in the $C^{k_0}(\mcNone\cup\mcNtwo)$-topology on the data.  For $r_1\le 0 $   let $(\whmcN{r_1}, \smet_{AB},\alpha)$ be as in the proof of  Proposition~\ref{P26XII22.1}.
We can choose the data to vary continuously in $C^{k_0}$ as $r_1$ varies. Cauchy stability guarantees that there exists $\epsilon_1>0$ so that the solution of the Cauchy problem with the data $\CSdata{\mcN,\infty}$ on $\mcN$ and the data $\CSdata{\whmcN{-\epsilon_1},\infty}$ on $\whmcN{-\epsilon_1}$ contains a future neighborhood of $\secN$.
A similar continuity argument applies to the data on the hypersurfaces $\whmcN{r_2}$ with $\epsilon_2\ge r_2>0$. One thus obtains a smooth vacuum spacetime metric as in Figure~\ref{F24XII22.1} with $r_1=-\epsilon_1$ and $r_2=\epsilon_2$.
\qedskip
\end{Remark}
}

A construction in the same spirit allows us to extend vacuum Cauchy data defined on a manifold with boundary $\Sigma$ beyond the boundary, to a larger initial data manifold $\check \Sigma$, where the boundary of $\Sigma$ becomes an interior hypersurface, while satisfying the vacuum general relativistic constraint equations:

\begin{Theorem}
 \label{T10III23.1}
 Let $\hyp$ be a manifold with boundary $\partial \hyp$ carrying smooth-up-to-boundary vacuum initial data $(g,K)$. There exists a manifold $\check\hyp$ with vacuum data $(\check g,\check K)$ and an isometric embedding of $(\hyp,g)$ into $(\check \hyp,\check g)$ such that $\check K$ coincides with $K$ on $\hyp$.
\end{Theorem}

\begin{figure}[t]
  \centering
\includegraphics[width=.65\textwidth]{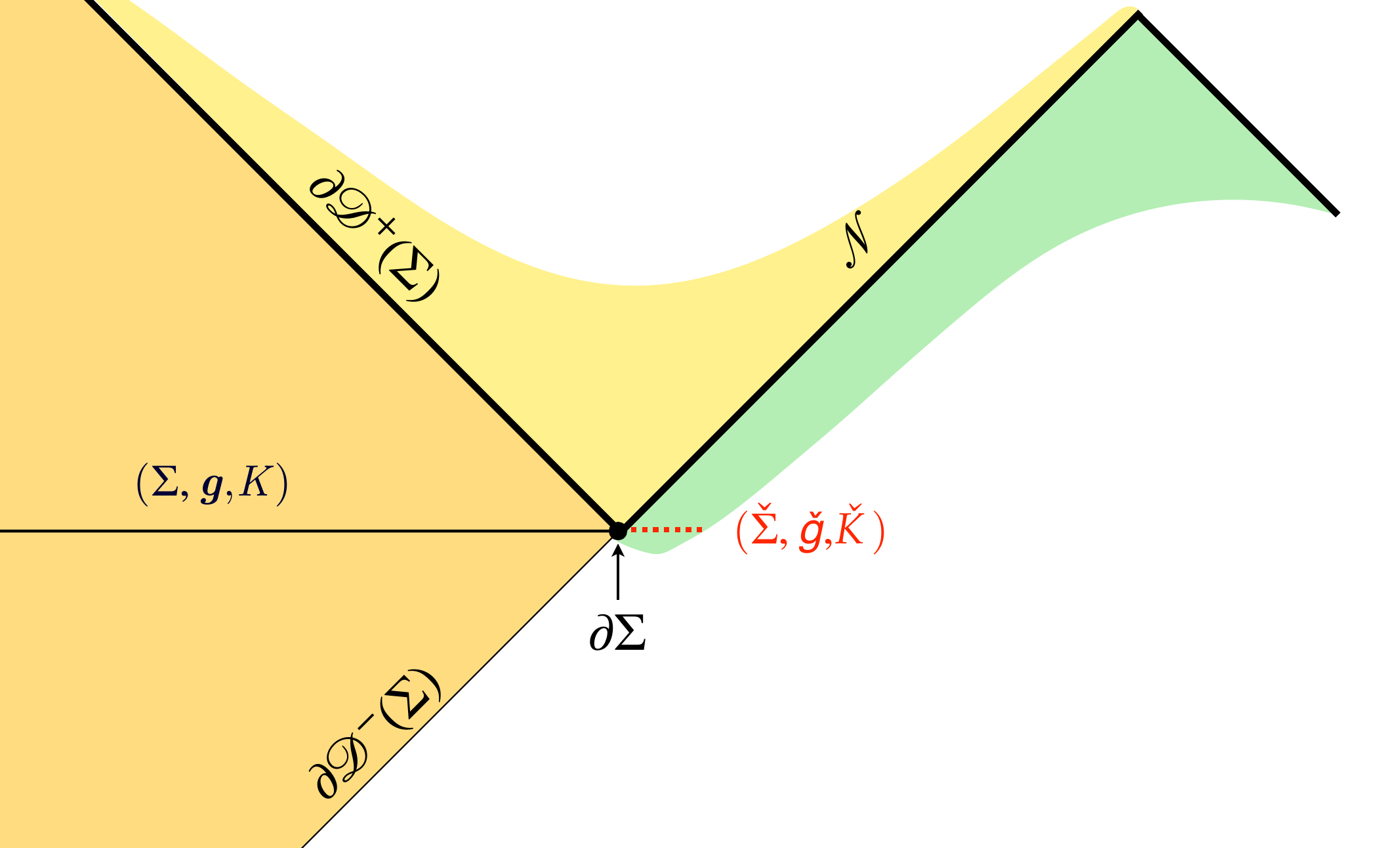}
\caption{Extending $\hyp$ to $\check \hyp$.
}
   \label{F4II23.1}
\end{figure}

\proof
The Cauchy data $(\hyp,g,K)$ induce spacelike vacuum data $\CSdata{\partial\hyp,\infty}$ on $\partial\hyp$.
The maximal globally hyperbolic development of $(\hyp, g,K)$ induces smooth characteristic data, compatible with $\CSdata{\partial\hyp,\infty}$, on  a null boundary emanating from $\partial \hyp$ in the direction of the null normal pointing towards $\hyp$, denoted by $\partial \mcD^+(\hyp)$ in Figure~\ref{F4II23.1}.
 Choose any characteristic data $\CSdata{\mcN,\infty}$, compatible with $\CSdata{\partial\hyp,\infty}$, on a hypersurface $\mcN$ intersecting $\partial \mcD^+(\hyp)$ transversally at $\partial \hyp$; $\mcN$ provides a smooth continuation, to the future, of the hypersurface
 $\partial \mcD^-(\hyp)$ of  Figure~\ref{F4II23.1}. The construction of the proof of Proposition~\ref{P26XII22.1} provides a vacuum metric, say $\fourg$, defined in a neighborhood $\mcO$ of $\mcN$. The spacelike hypersurface $\hyp$ can be extended smoothly within $\mcO$ across $\partial \hyp$  to a spacelike hypersurface $\check \hyp$. The data induced on $\check \hyp$ by $\fourg$ provide the desired extension $(\check \hyp,\check g,\check K)$.
\qedskip

We can also find vacuum metrics which extend vacuum metrics on solid light cones. For this we need to truncate the cone at finite distance by a spacelike acausal hypersurface $\hyp$. Consider, then, smooth characteristic vacuum data  
on a light cone $\mcC_p$ with vertex at $p$.
By~\cite{ChruscielSigma} there exists a  neighborhood $\mcO$ of $p$ and a smooth vacuum metric  $\fourg$ defined on $\mcO\cap
 J^+(p)$ which realises the data. (In space-time dimension  four  the set $\mcO$ constitutes a full future neighborhood of $\mcC_p$.) Let $\hyp\subset \mcO$ be any smooth spacelike hypersurface included in $J^+(p)$ with smooth compact boundary on $\mcC_p$:
$$
 \secN:= \partial\hyp = \hyp \cap \mcC_p  \,.
$$
We denote by $\mcC_p^{\secN}$ the cone $\mcC_p $ truncated at $\secN$,
\begin{equation}\label{28V23.1}
 \mcC_p^{\secN}:= \mcC_p\cap J^-(\secN)
 \,,
\end{equation}
 see Figure~\ref{F5II23.1}. We have:
\begin{figure}
  \centering
\includegraphics[width=.6\textwidth]{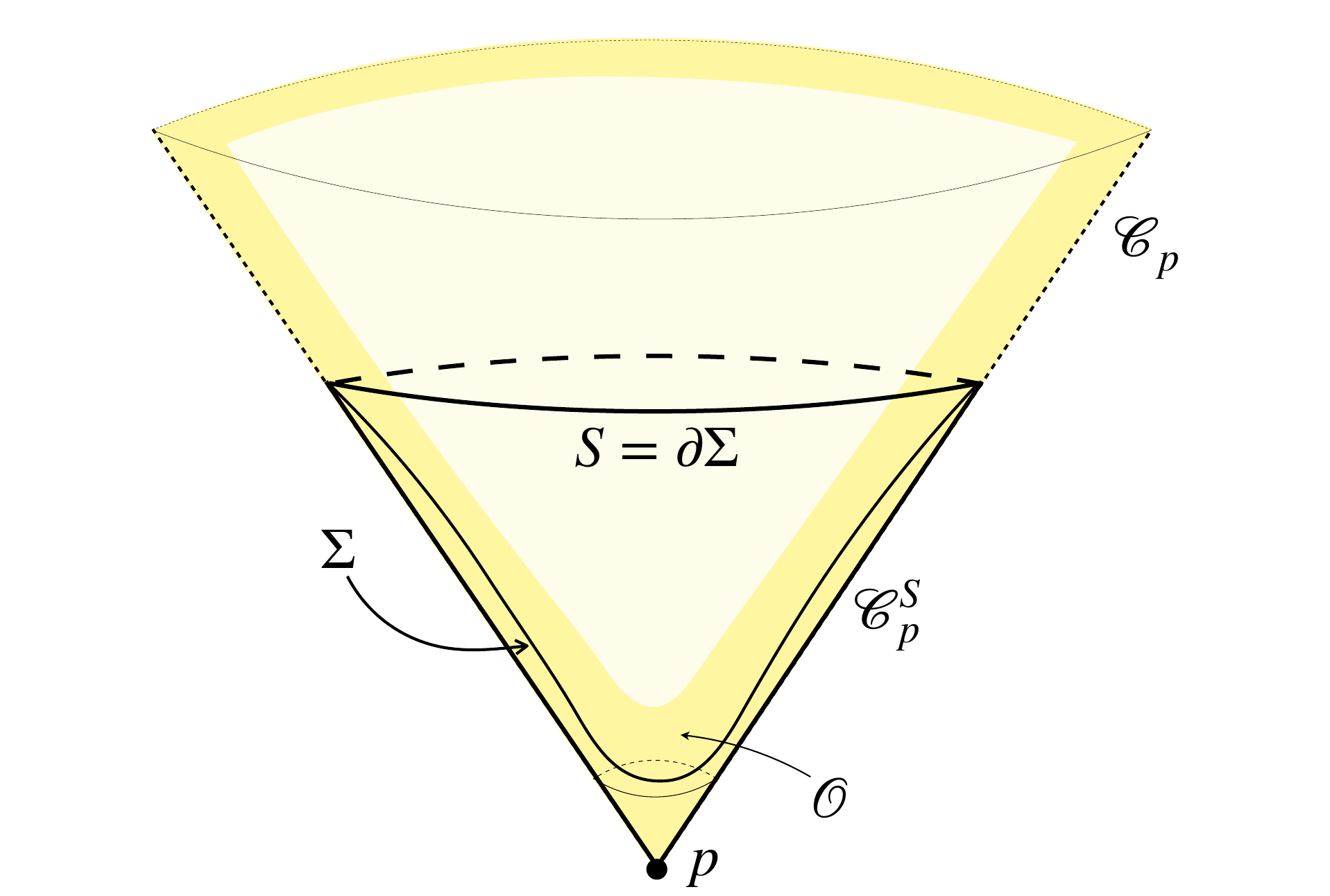}
\caption{Truncating the light cone of $p$ at $\partial\hyp$.
}
   \label{F5II23.1}
\end{figure}

\begin{figure}
  \centering
\includegraphics[width=.6\textwidth]{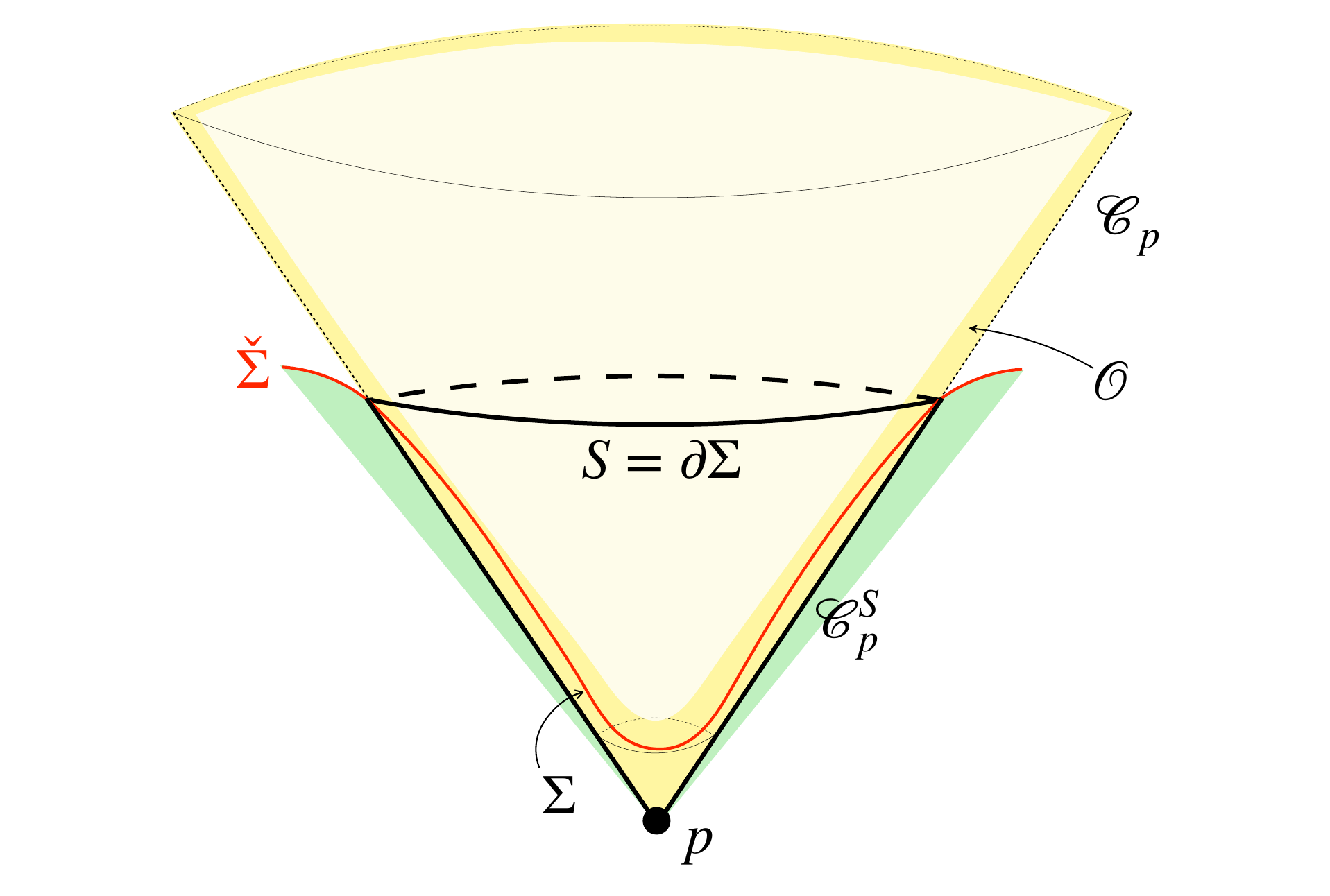}
\caption{Extending a vacuum metric on a truncated future cone $J^+( p)\cap J^-(\hyp)$ to a neighborhood  of
$(J^+( p)\cap J^-(\hyp))\setminus \{p\}$.
}
   \label{F5II23.2a}
\end{figure}

\begin{proposition}
 \label{P4II23.2}
For any smooth characteristic vacuum data
 on a truncated light cone $\mcC_p^{ \secN}$ as above, there exists a smooth vacuum metric  realising the data defined in a neighborhood of {
 $$\mcC_p^{ \secN}\setminus \{p\}
 \,,
 $$
see Figure~\ref{F5II23.2a}.
 }
\end{proposition}

\proof
By Theorem~\ref{T10III23.1} we can extend $\hyp$ beyond its boundary to a new spacelike hypersurface $\check\hyp$ as in Figure~\ref{F5II23.2a}.
 Solving backwards in time the Cauchy problem with the extended data one obtains a vacuum metric defined in a neighborhood $\mcO$ of $\mcC_p^{ \secN}\setminus\{p\}$.
 \qedskip

\begin{figure}
  \centering
\includegraphics[width=.6\textwidth]{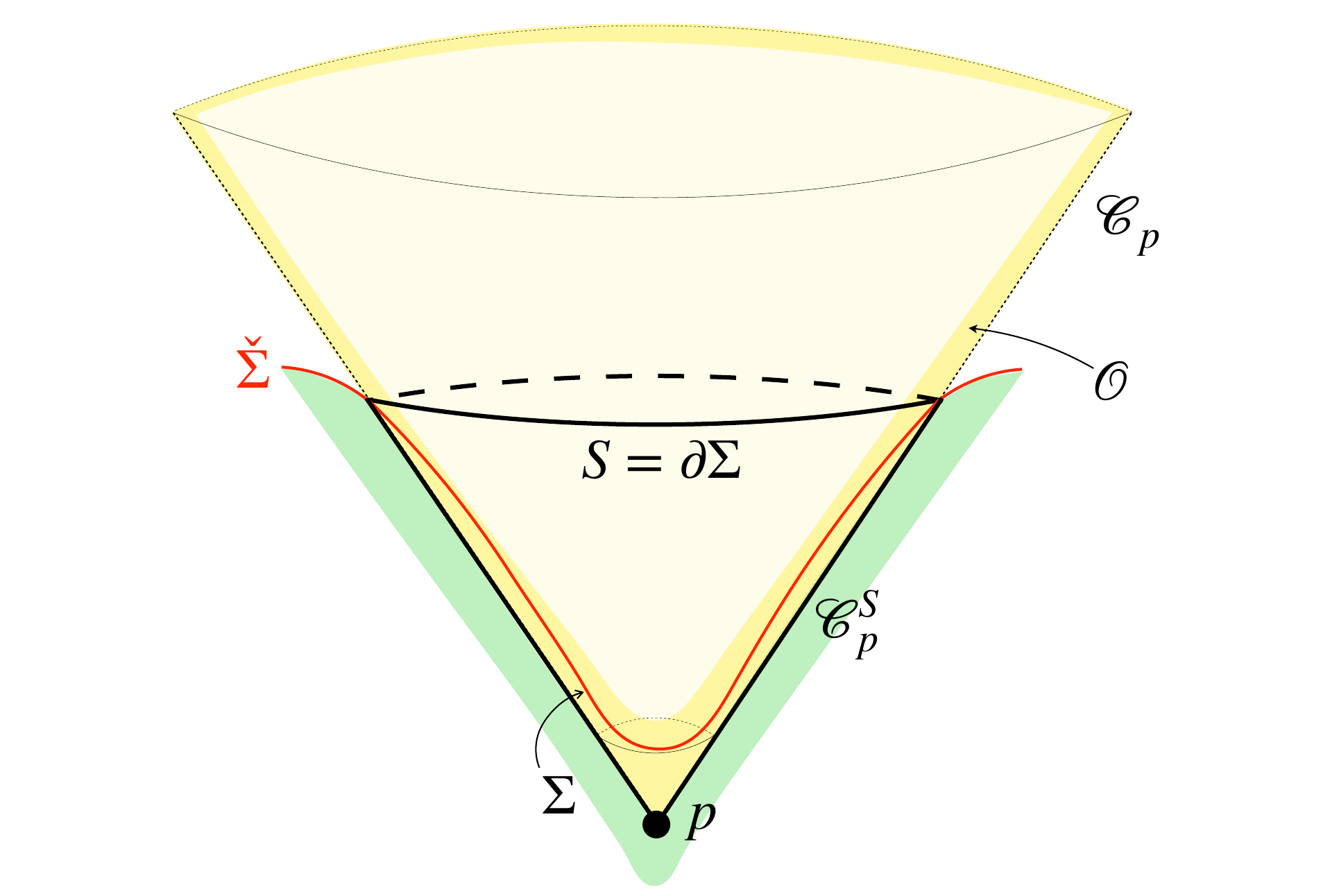}
\caption{Extending a vacuum metric on a truncated future cone $J^+( p)\cap J^-(\hyp)$ to a neighborhood thereof.
}
   \label{F5II23.2b}
\end{figure}

The question then arises, whether we can always obtain a full neighborhood of $\mcC_p^{ \secN}$, as in Figure~\ref{F5II23.2b}.
The problem is, that the domain of existence of $\fourg$ might be shrinking as $p$ is approached, as
seen in Figure~\ref{F5II23.2a} and  made clear by the following considerations:

By definition of smooth characteristic data near the tip  of a light-cone,
there exists a smooth Lorentzian metric inducing the data. After solving the Einstein equations to the future of the light-cone as in~\cite{ChruscielSigma} we obtain a smooth metric, say
$\widehat\fourg$, defined in a neighborhood of $p$ which coincides with $\fourg$ in $J^+(p)$, and thus is vacuum there. But we have no reason to expect that it will coincide with $\fourg$  away from $J^+(p)$, nor that it will be vacuum away from $J^+(p)$.

As an attempt to address this issue, we will use $\widehat\fourg$-normal coordinates near $p$ to study  the behaviour  of $\fourg$ there, keeping in mind that $\widehat\fourg$ extends smoothly $\fourg|_{\mcC_p}$ in a neighborhood of $\mcC_p^{ \secN}\setminus \{p\}$; these coordinates are the only reason why we need the metric $\widehat \fourg$.

Let, thus, $(t,\vec x) \equiv (t,x^i)\equiv (x^\mu)$ be normal coordinates centred at $p$ for the metric $\widehat\fourg$, in these coordinates the light-cone is given by the equation $t=|\vec x|$, and there exists a constant $C$ such that for $ |\vec x|\le 2 t$ we have
\begin{equation}\label{26V23.1a}
  |\widehat\fourg_{\mu\nu} - \eta_{\mu\nu}| \le C (t^2 + |\vec x|^2) \le 5 C t^2\,,
  \quad
  |\partial_\sigma \widehat\fourg_{\mu\nu}  | \le C  t
  \,.
\end{equation}
For any $k \ge 2$  for $2\le i\le k$ it holds that
\begin{equation}\label{26V23.2a}
  |\partial_{\sigma_1} \cdots \partial_{\sigma_i} \widehat\fourg_{\mu\nu}  | \le C
  \,,
\end{equation}
where the constant might depend upon $k$.
In what follows we choose some $k> n/2 +1$, to guarantee that the solutions of the
spacelike Cauchy problem for the Einstein equations with data in $(g,K)\in H_{k+1}\times H_k$ are in $C^2$.

Let
\begin{equation}\label{28V23.14}
 \Sigma_c[x]=\{t=c\}
 \,.
\end{equation}By Cauchy stability, for $t>0$ the intersection $\Sigma_t\cap \mcO$ of the domain $\mcO$ of definition of the vacuum metric $\fourg$ contains the set
\begin{equation}\label{26V23.1}
  \{ |\vec x| \le f(t) \}\,,
  \ \mbox{for some function satisfying $f(t)>t$.}
\end{equation}
Replacing $f$ by a smaller function if necessary,  we can assume that for $t>0$ we have
$$
 t < f(t)\le 2t
 \,.
$$
Passing to a smaller function $f$ again if necessary, smoothness of $\fourg$ implies that on the set $\{|\vec x|\le f(t)\}$ we will have
\begin{equation}\label{26V23.1ax}
  | \fourg_{\mu\nu} - \eta_{\mu\nu}|   \le 10 C t^2\,,
  \quad
  |\partial_\sigma  \fourg_{\mu\nu}  | \le 2C  t
  \,,
\end{equation}
as well as, for $2\le \ell \le k$ ,
\begin{equation}\label{26V23.2ax}
  |\partial_{\sigma_1} \cdots \partial_{\sigma_\ell} \fourg_{\mu\nu}  | \le 2 C
  \,.
\end{equation}

Finally, again making $f$ smaller if necessary we can assume that the function $f$ is continuous and increasing.

For small $s>0$, say $s\le s_0<1/2$ for some $s_0$ smaller than the injectivity radius, consider the   \emph{scaling map}
\begin{equation}\label{30V23.1}
 (\tau, \vec y) \equiv (y^0,\vec y) \equiv  (y^\mu)
   \mapsto  (t,\vec x) \equiv  (x^\mu) := (s y^\mu) \,.
\end{equation}
Using \eqref{30V23.1} we obtain a family of scaled  metrics, solutions of vacuum Einstein equations with cosmological constant $ \Lambda_s:=s^2 \Lambda$:
\begin{equation}\label{26V23/4}
  \fourg[s]_{\mu\nu}(y^\alpha)= \fourg_{\mu\nu}(s y^\alpha)
  \,.
\end{equation}
Set
$$ \Sigma_c [y]:= \{\tau=c\}
\,,
$$
then on $\Sigma_1[y]$ the metric $\fourg [s]$ is defined on a set containing the coordinate ball
$$
 B[s]:= \{|\vec y|\le f(s)/s\}
 \,,
 \
 \mbox{with the radius satisfying $1< f(s)/s \le 2$.}
$$
Let $(B[s],g[s],K[s])$ be the Cauchy data induced by $\fourg[s]$ on
$$
 \{
  \tau=1,|\vec y|\le f(s)/s
   \}
    \subset \Sigma_1 [y]
 \,.
$$
It follows from \eqref{26V23.1ax}-\eqref{26V23.2ax}  that for $\vec y \in B[s]$ we have
\begin{equation}\label{26V23.1b}
  |g[s]_{ij}-\delta_{ij}| \le 10 C s^2  \,,
  \quad
  |\partial_{y^{\ell}} g[s]_{ij}  | \le 2 C s^2  \,,
  \quad
  |K[s]_{ij}  | \le C'  s ^2
  \,,
\end{equation}
for some constant $C'$, and that for $2\le \ell \le k$ it holds that
\begin{equation}\label{26V23.2b}
  |\partial_{y^{i_1}} \cdots \partial_{y^{i_\ell}}  g_{ij} | \le 2  C  s^\ell    \le 2 C s^2
  \,.
\end{equation}
Hence the Cauchy data set $(B[s],g[s],K[s])$ tends,   in $H_{k+1}(B[s])\times H_{k}(B[s])$ norm, as $s$ tends to zero, to the Minkowskian one, $(\mcB(1),\delta_{ij},0)$, where $\mcB(1)$ is the unit coordinate ball centered at the origin in $\R^n$. Standard hyperbolic estimates imply that

\begin{enumerate}
  \item the boundary 
  of the maximal past globally hyperbolic development of the data $(B[s],g[s],K[s])$ is generated by null geodesics normal to $\partial B[s]$, and
  \item  on the past domain of dependence of the data, say $\mcD^-[s]$ we have
\begin{equation}\label{28V23.11}
  |\partial_{y^{\alpha_1}}\cdots \partial_{y^{\alpha_\ell}}
   \big(\fourg[s]_{\mu\nu}-\eta_{\mu\nu}\big)| \le \check C s^2
  \,,
  \end{equation}
  for some constant $\check C$.

  \item
Furthermore,  the maximal past globally hyperbolic development of $(B[s],g[s],K[s])$ approaches the Minkowskian past domain of dependence as $s\to0$ in the sense that is made clear by the following: every  generator $x^\mu(t)$ of  $\partial \mcD^-[s]$ starting at a point $p\in S^{n-1}\subset \partial B[s]$ lies a distance not further than
\begin{equation}\label{1VI22.1}
  \tilde Cs^2
  \,,
\end{equation}
for some constant $\tilde C$, from the generator of the Minkowskian past domain of dependence of $\{\tau=1, \vec y \in B[s]\}$ issued from the same point $p$ on $\partial B[s]$. Therefore
\begin{equation}\label{2VI23.11}
  \partial \mcD[s] \subset \{|\vec y| \ge s^{-1}f(s) -1 - \tilde C s^2
  \}
  \,.
\end{equation}
\end{enumerate}

\begin{figure}
  \centering
\includegraphics[width=.5\textwidth]{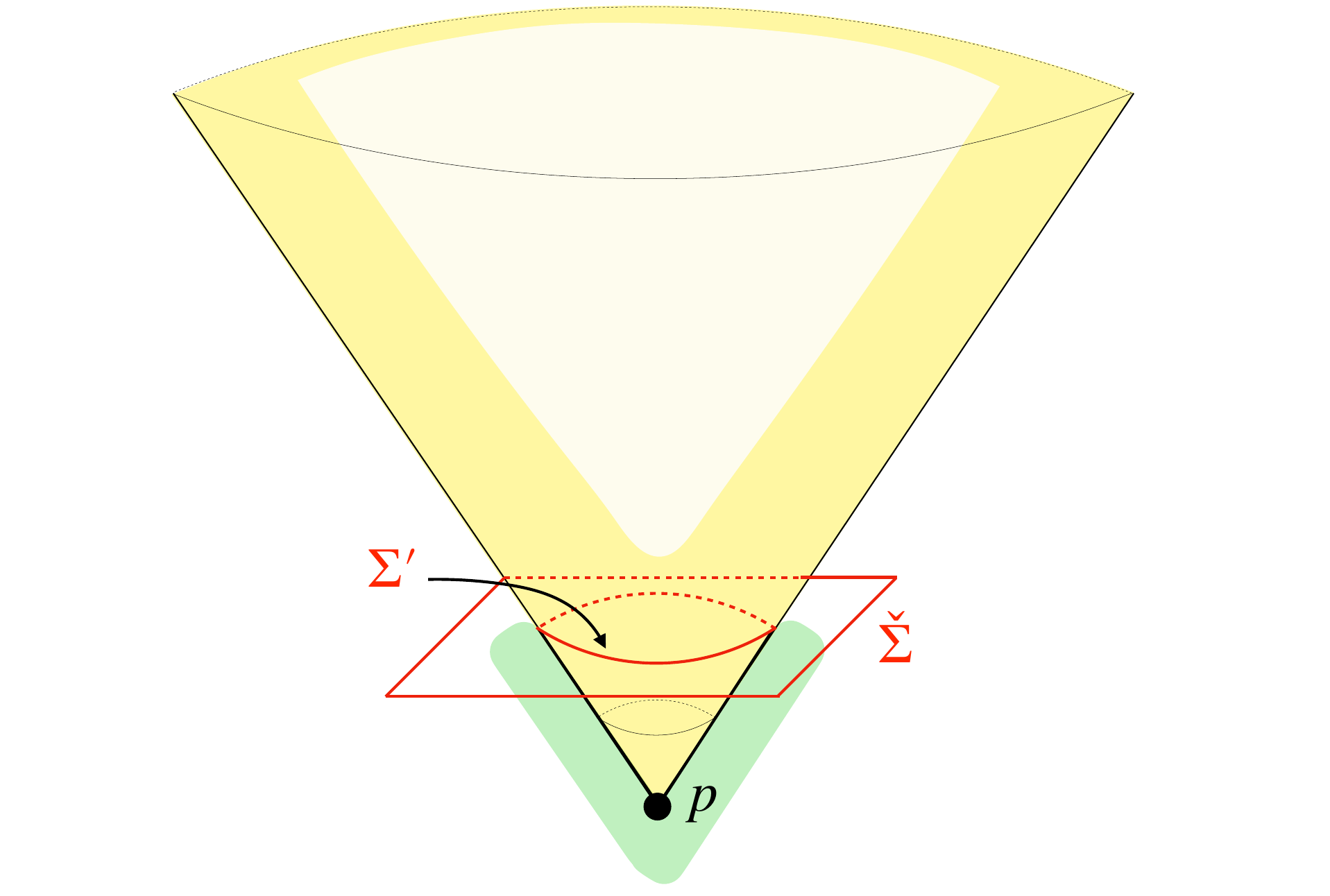}
\caption{Extending a vacuum metric on a truncated future cone $J^+( p)\cap J^-(\hyp')$ to a neighborhood  of
$J^+( p)\cap J^-(\hyp') $.
}
   \label{F4IV23.1}
\end{figure}

Now, in the scaled-back original coordinates  the Minkowskian past domain of dependence of the set
$$
 \{t=s, |\vec x|\le f(s)\}
$$
is a truncated solid cone with vertex at $ (s- f(s) ,\vec 0)$, and note that $s- f(s) <0$.
Next, it follows from \eqref{2VI23.11} that the boundary of the set $\{t=0\}\cap \mcD^-[s]$ lies inside the set
$$
 \{|\vec x| \ge f(s) -s -\tilde C s^3\} \,.
$$
Hence $\{t=0\}\cap \mcD^-[s]$ will contain a neighborhood of the origin whenever
\begin{equation}\label{28V23.21x}
 \liminf_{s\to0} \frac{f(s) -s }{s^{ 3}} = \infty
  \,.
\end{equation}
Whether or not \eqref{28V23.21x} holds in general is not clear.
However, we claim
that \eqref{28V23.21x} is satisfied if the truncating section $\secN$ of $\dot J^+(p)$   in \eqref{28V23.1} is close enough to $p$:

\begin{proposition}
 \label{P4II23.2a}
 If $\secN$ is close enough to $p$,   there exists a  metric as in Proposition~\ref{P4II23.2} which is defined in a neighborhood of
 $\mcC_p^{ \secN}
 $ as in Figure~\ref{F4IV23.1}.
\end{proposition}

\proof
We continue to use $\check \fourg$-normal coordinates.
%
We can carry out the hand-crank construction as in Figure~\ref{F4II23.1}, with $\Sigma$ there being the unit $\vec y$--coordinates ball within $\Sigma_1[y]$, and
with $\mcN$ in Figure~\ref{F4II23.1} being the part of $\dot J^+(0)$ between $\Sigma_1[y]$ and $\Sigma_2[y]$, with the transverse free data choosen to tend to the Minkowskian ones there as $O(s^2)$   in any finite Sobolev norm. Given any $k_1\in \N$ we can find  $k> k_1$ in \eqref{26V23.2a} large enough so that the extended solution on the green region of Figure~\ref{F4II23.1} tends to the Minkowski metric there,  in $C^{k_1}$  norm, as
$O(s^2)$. This shows that for small enough $s$, say $s\le \check s$, the vacuum initial data on the unit $\vec y$--coordinate ball within $\Sigma_1[y]$
 can be extended to  $\{\tau=1, |\vec y|\le 2\}$. It follows that the function $f(s)$ in \eqref{2VI23.11} can be chosen  to be $2s$, so that \eqref{28V23.21x} holds. One  obtains a spacetime as in Figure~\ref{F4IV23.1} with  $\check \Sigma = \Sigma_{\check s}[x]$ and $\Sigma'= \{t=\check s\,,\, |\vec x|\le \check s\}$.
\qedskip

As a corollary we obtain:

\begin{Corollary}
 \label{C4II23.2}
 Let $k\in \N\cup\{\infty\}$.
For any vacuum data  $\CSdata{\{p\},k}$ at a point $p$, there exists a vacuum metric defined in a neighborhood of $p$  realising the data.
\end{Corollary}

\proof
By definition, there exists
a smooth Lorentzian metric $\mathring \fourg$ inducing $\CSdata{\{p\},k}$; we emphasise that we do not assume that $\mathring \fourg$  is vacuum. Denote by $ J^+(p; \mathring \fourg)$  the causal future of $p$ in $\mathring \fourg$, and by
$\mcC_p $   the light cone of $\mathring \fourg$ emanating from $p$. Let $\mathring g_{AB}dx^Adx^B$ be the tensor field of signature $(0,+,\ldots,+)$ obtained by restricting $\mathring \fourg$ on $\mcC_p $. By~\cite{ChruscielSigma} there exists a  neighborhood $\mcO$ of $p$ and a smooth vacuum metric $\fourg$ defined on
$$\mcO\cap
 J^+(p; \mathring \fourg) = \mcO\cap
 J^+(p;  \fourg)
  \,,
$$
with
$\mcC_p$ being the light cone of $\fourg$, and with $\fourg$
 inducing on $\mcC_p$ the same degenerate tensor as $\mathring \fourg$.
 It follows that the data induced at $p$ by $\fourg$ coincide with those induced by $\mathring \fourg$, i.e.\ $\CSdata{\{p\},k}$. The result follows by Proposition~\ref{P4II23.2a}.
\qed

\section{The ``Fledermaus  construction''}
 \label{s22XII22.1}

We have shown in Section~\ref{s26XII22.1} how to find a vacuum metric which realises vacuum characteristic data $\CSdata{\mcN,k}$ on a hypersurface as an \emph{interior submanifold}. Here we describe a construction which realises vacuum characteristic data on two transverse vacuum characteristic hypersurfaces $\mcNone\cup \mcNtwo$ as an \emph{interior submanifold with corner} in a vacuum spacetime. This should be contrasted with Theorem~\ref{T5I23.1}, which realises the data as the \emph{boundary} of a spacetime with boundary-with-corner. Not unexpectedly, the resulting metric is only uniquely defined to the future of $\mcNone\cup\mcNtwo$.

Indeed, in this section we use a ``Fledermaus  construction'' to show:

\begin{Proposition}
 \label{P22XII22.1}
Consider a smooth vacuum  initial data on  two hypersurfaces
$$\mcNonep\approx [0,1]\times \secN \ \mbox{and} \ \mcNtwop\approx [0,1]\times \secN
$$
meeting transversally at a compact submanifold $\secN$.
There exists a smooth solution of vacuum Einstein equations $\fourg$ which is defined in a neighbourhood of $\mcNonep\cup\mcNtwop$ and which realises the data.
\end{Proposition}

\begin{remark}
 \label{R31I23}
{\rm
The metric constructed in Proposition~\ref{P22XII22.1} is uniquely determined by the characteristic data on the hypersurface $\hmcN{}_{[0,1]}\cup\mcNtwop\cup\mcNonep\cup\underline{\hmcN{}}_{[0,1]}$ of Figure~\ref{F22XII22.1}.
\begin{figure}[t]
  \centering
\includegraphics[width=.8\textwidth]{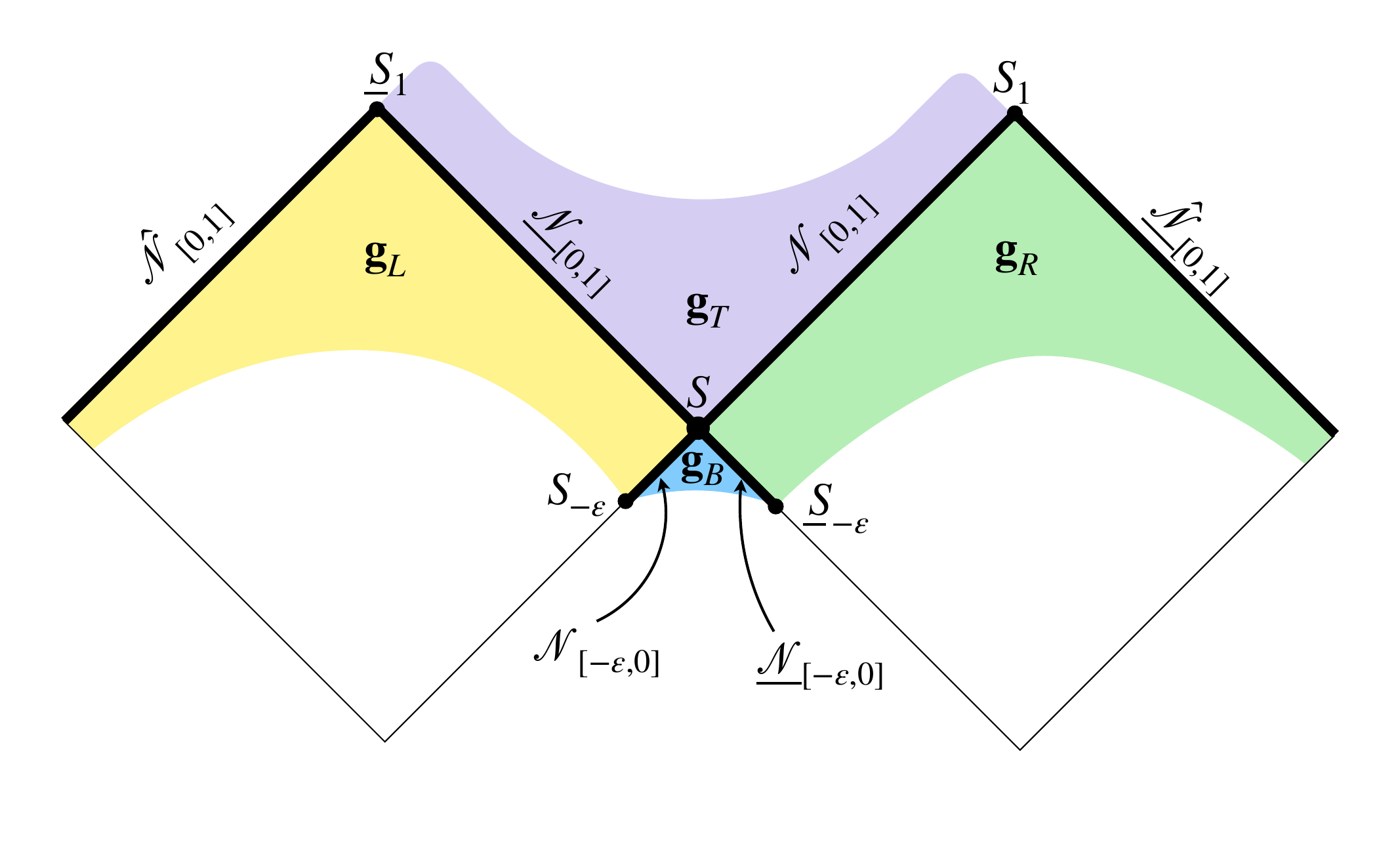}
\caption{The ``Fledermaus  construction'' to embed characteristic initial data on two transverse hypersurfaces in a vacuum spacetime.
}
   \label{F22XII22.1}
\end{figure}
 }
 \qed
 \end{remark}

\proof
Let us denote by $\fourg_T$,
where ``$T$'' stands for ``top'',
the smooth solution of the vacuum Einstein equations obtained by solving the characteristic Cauchy problem to the future of $\mcNonep\cup\mcNtwop$ with the given data. The solution induces  a set of spacelike vacuum data $\CSdata{\secN,k}$ with $k=\infty$ on $\secN$ and sets of characteristic vacuum data  $\CSdata{\mcNonep,k}$ and  $\CSdata{\mcNtwop,k}$ on $\mcNonep $ and  $ \mcNtwop $, again with $k=\infty$.

We view $\mcNonep$ as a subset of a smooth hypersurface
$$\mcN=[-1,1]\times \secN
 \,,
$$
and we
view $\secN$ as the subset $\{r=0\}$ of  $\mcNone$. We denote by $\secN_{r}$ the crossection $\{r\}\times \secN$.
For $r\ge 0$ the metric $\fourg_T$ induces smooth spacelike vacuum data $\CSdata{\usecN_r,k}$ with $k=\infty$ on $\secN_r$.

Similarly we view  $\mcNtwop$ as a subset of a smooth hypersurface
$$\mcNtwo=[-1,1]\times \secN
 \,,
$$
with crossections $\{r\}\times \secN\subset \mcNtwo$ denoted by $\underline{\secN}_{r}$, and with induced vacuum data $\CSdata{\usecN_r,k}$ for $r>0$.

Let
$$\hmcN{}_{[0,1]}:= [0,1]\times \usecN_{1}
$$
be a null hypersurface meeting $\mcNtwo$ transversally at $\usecN_1$ towards the past; see Figure~\ref{F22XII22.1}.
 We choose any Isenberg-Moncrief
 fields $(\smet_{AB},\alpha)$
  on $\hmcN{}_{[0,1]}$ compatible with $\CSdata{\usecN_1,k}$ and we solve the characteristic Cauchy problem to the past with data on $\hmcN{}_{[0,1]}\cup \mcNtwop$.
 One thus obtains a vacuum metric, say $\fourg_L$, where $L$ stands for ``left'', on the left wing of the Fledermaus . Uniqueness of solutions of transport equations for the transverse derivatives of the metric along $\mcNtwop$ implies that the metric $\fourg_L$ extends smoothly $\fourg_T$ across $\mcNtwop$.
 The intersection of the domain of existence of $\fourg_L$ with $\mcNonem$ contains the hypersurface
 $$\mcNone_{[-\epsilon,0]}:=[-\epsilon,0]\times \secN\subset \mcNone
  \,,
 $$
which will be made-use of shortly.

 A similar construction provides a smooth vacuum metric $\fourg_R$ on the right wing of the Fledermaus, extending smoothly $\fourg_T$ across $\mcNtwom$, with domain of existence   containing a hypersurface %
 $$
  \mcNtwo_{[-\epsilon,0]}:=[-\epsilon,0]\times \secN\subset \mcNtwo
  \,.
  $$
  The three metrics $\fourg_L$, $\fourg_T$ and $\fourg_R$ match smoothly at $\mcNonep\cup\mcNtwop$, in particular also at $\secN$.

Let $\fourg_B$ be obtained by solving the characteristic Cauchy problem to the past with data on $\mcNone_{[-\epsilon,0]}\cup\mcNtwo_{[-\epsilon,0]} $. From what has been said so far it should be clear that the four metrics $\fourg_L$, $\fourg_T$, $\fourg_R$ and $\fourg_B$ match smoothly wherever more than one is defined, and provide the desired smooth vacuum metric $\fourg$ defined on a neighborhood of $\mcNonep\cup\mcNtwop$.
\qed

\section{Null hypersurfaces and spacelike gluing}
 \label{s26XII22.3}

It is well known~\cite{Moncrief75,Corvino} that the existence of Killing vectors near a spacelike Cauchy surface provides an obstruction to the Corvino-Schoen approach to spacelike gluing; see, however, \cite{CzimekRodnianski}. More precisely, there is an obstruction to the gluing construction based on the implicit function theorem involving the adjoint of the linearised constraint operator. We show in Appendix~\ref{CWs23II22.1}  that an identical obstruction arises in the characteristic gluing.

In fact, the obstruction arising from Killing vectors is a local one:
We shall say that there are \emph{no local Killing vectors near $\secN$} if  the Killing vector equation has only trivial solutions on all sufficiently small neighborhoods of $\secN$. Formally: every neighborhood of $\secN$ contains another neighborhood of $\secN$ on which only trivial solutions of the Killing equations exist.

It might be of some interest to note that, given a submanifold $\subman$ of $\mcM$ of any dimension and type, the  notion of \emph{absence of local Killing vectors}  can be defined in terms of submanifold data $\CSdata{\subman,k}$ of order $k\ge 1$,  For this, note that the Killing equations at $\subman$,
\begin{equation}\label{5II23.1}
 (\nabla_\mu X_\nu + \nabla_\nu X_\mu) |_\subman = 0
 \,,
 \end{equation}
and their derivatives in both transverse and tangential directions to $\subman$, up to order $k$,
evaluated at $\subman$
(e.g., \eqref{5II23.1} together with
$$
  \nabla_\mu\nabla_\nu X_\rho  |_\subman = R_{\sigma \mu\nu\rho} X^\sigma
$$
when $k=2$), can be viewed as an overdetermined set of equations for the jets of order $k$ over $\subman$ of a vector field $X$. We will say that \emph{there are no local Killing vectors at $\subman$} if there exists $k\ge 1$ such that these equations have only the trivial solution. Standard arguments show that the absence of local Killing vectors \underline{at} $\subman$  implies that every metric near $\subman$ compatible with $\CSdata{\subman,k}$ will have no local Killing vectors \underline{near} $\subman$.

\begin{figure}
  \centering
\includegraphics[width=.7\textwidth]{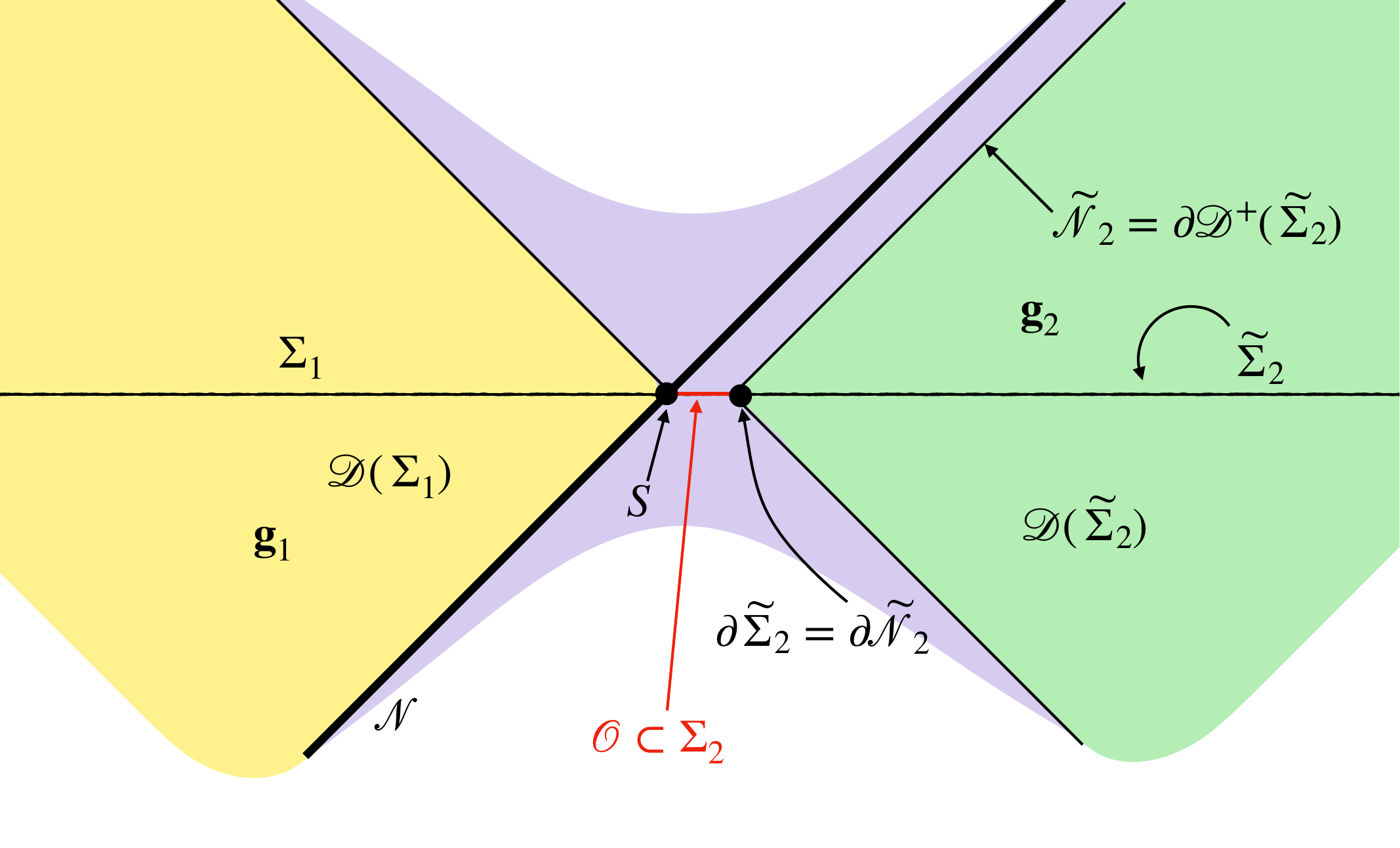}
\caption{After the gluing, zoom to the gluing region; $\tilde \hyp_2$ is $\hyp_2\setminus \mcO$.
}
   \label{F17XII22.2c}
\end{figure}

We have:

\begin{theorem}
  \label{T18XII22.1}
  Consider two smooth vacuum metrics $\fourg_1$ and $\fourg_2$ on $\mcM$  and let $\mcN\subset \mcM$ be a hypersurface in $\mcM$ which is null both for $\fourg_1$ and $\fourg_2$. Let $\secN\subset \mcN$ be a  compact
  cross-section of $\mcN$ and suppose that there are no local Killing vectors near $\secN$ for $\fourg_1$.
  If $\fourg_1$ and $\fourg_2$ are sufficiently close to each other near $\secN$ in $C^5$-topology, then
  there exists a smooth vacuum metric $\fourg$ on $\mcM$ and a null hypersurface $\tmcN2$
   in $\mcM$ with spacelike boundary $\partial\tmcN2$, with  $\tmcN2$ near to
   $\mcN \setminus J^-(\secN)$ (cf.\ Figure~\ref{F17XII22.2c}), so that
   \begin{enumerate}
     \item $\fourg$ coincides with $\fourg_1$ on
      $\overline{J^+(\mcN)\setminus J^+(\secN)}$,
       and
     \item $\fourg$ coincides with $\fourg_2$ on
      $\overline{J^-(\tmcN2)\setminus J^-(\partial\tmcN2)}$.
   \end{enumerate}

   In particular $\fourg$  induces the original vacuum data $\noCdata{\mcN\cap J^-(\secN),k}$ induced by $\fourg_1$ on $\mcN\cap J^-(\secN)$ for any $k\in \N$, with the data $\noCdata{\mcN\cap J^+(\secN),k}$ induced by $\fourg$ on $ \mcN\cap J^+(\secN) $ being close to the data induced by $\fourg_2$ there,
    and with the data $\noCdata{\tmcN{},k}$ induced by $\fourg$ on $ \tmcN{}  $ coinciding with the data induced by $\fourg_2$ there.
\end{theorem}

\begin{figure}
  \centering
\includegraphics[width=.7\textwidth]{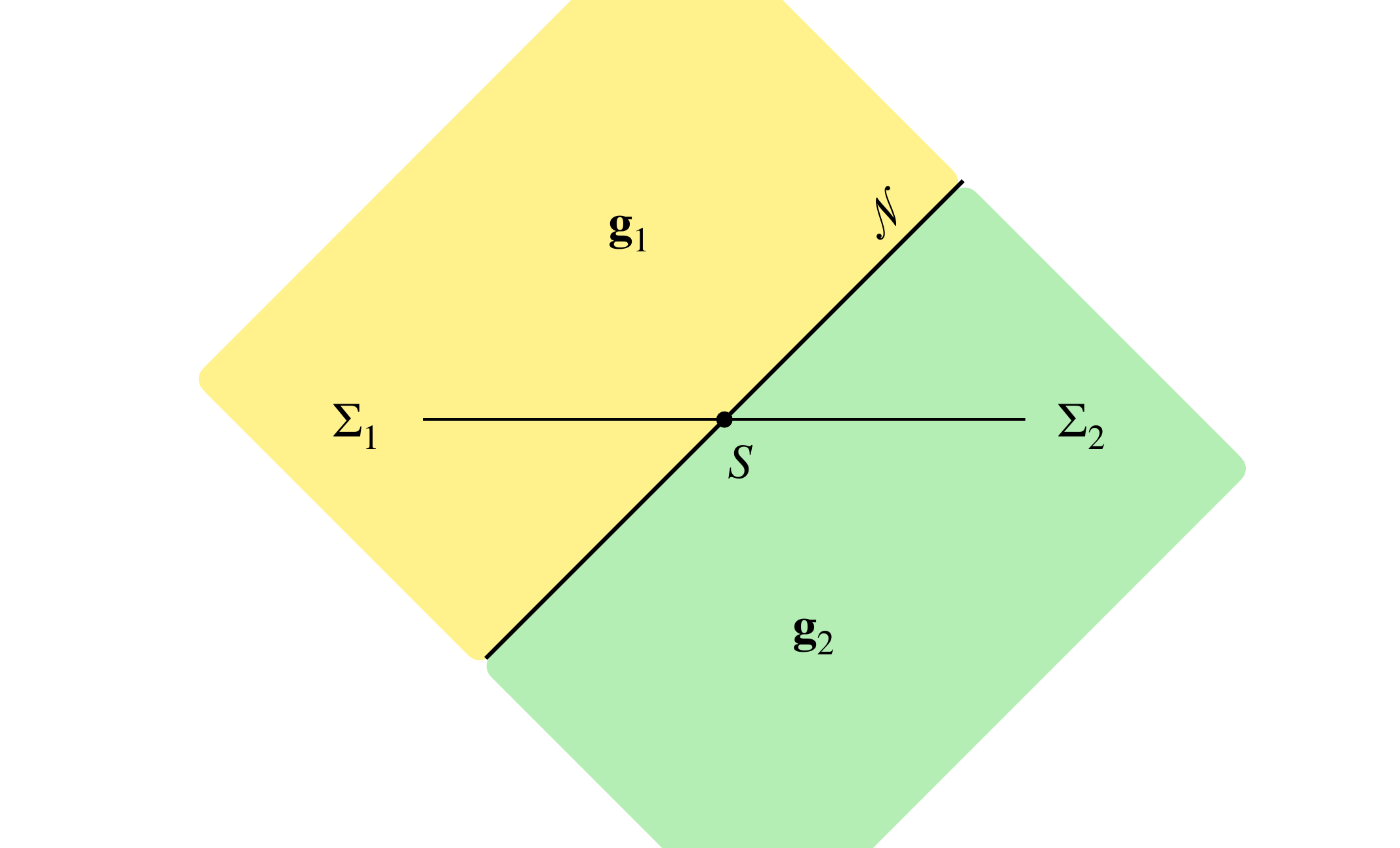}
\caption{Before the gluing of metrics with nearby characteristic data along a null hypersurface $\mcN$.
}
   \label{F22I23.1}
\end{figure}

\begin{remark}
 \label{R22I23.3}
 {\rm
In Theorem~\ref{T18XII22.1} it suffices to assume that the metric $\fourg_1$ is defined to the future of $\mcN$ and $\fourg_2$ is defined to the past of $\mcN$, as in Figure~\ref{F22I23.1}.
}
\qed
\end{remark}

\begin{remark}
 \label{R22I23.2}
{\rm
In the spacelike gluing the obstruction arising from Killing vectors can be circumvented by gluing to a family of initial data which carries a set of compensating parameters;
the gluing construction chooses a member of the family. There is an obvious version of Theorem~\ref{T18XII22.1} in such a situation, whenever a family of metrics with compensating parameters is available;
compare Section~\ref{s4II23.1} below.
 In particular one has a similar result for  gluing a vacuum metric $\fourg _1$ with a member $\fourg_2$ of the Kerr,
Kerr-de Sitter,  or Kerr Anti-de Sitter family. Note that  the condition of nearness to a member of the Kerr-(A)dS family is more severe, as compared to the $\Lambda=0$ case, in the following sense:
nearness to Kerr can be achieved by receding in spacelike directions for a large class of asymptotically Minkowskian initial data sets, while no such construction is known when $\Lambda\ne 0$. Compare~\cite{ChDelayAH,CortierKdS,HintzdSBH}.
}
\qed
\end{remark}

\proof
We can choose  spacelike hypersurfaces $\hyp_1$ and $\hyp_2$  as in Figure~\ref{F22I23.1} so that the vacuum Cauchy data  $(\hyp_1,g_1,K_1)$ and $(\hyp_2,g_2,K_2)$, induced by the respective spacetime metrics $\fourg_1$ and $\fourg_2$, are near to each other in a neighborhood of $\secN$ in a $C^5\oplus C^4$ topology. By \cite[Section~8.6]{ChDelay}
the data $(\hyp_1,g_1,K_1)$ and $(\hyp_2,g_2,K_2)$ can be smoothly glued together to a smooth vacuum data set
$(\hyp_1\cup\hyp_2,g ,K )$, so that $(g,K)$ coincides with the original Cauchy data except for a small neighborhood $\mcO\subset \hyp_2$ of $\secN$. Solving the Cauchy problem with these data one obtains the desired spacetime, see Figure~\ref{F17XII22.2c}. The hypersurface
$\tmcN2$ is taken to be $\partial \mcD^+(\thyp2)$, where $\thyp2= \hyp_2\setminus \mcO$.
\qedskip

\section{Gluing cross-section data to  Kerr data}
\label{s4II23.1}

We turn our attention now to the question addressed in~\cite{ACR1},  of gluing two sets of cross-section data, one of them arising from the Kerr family.
For definiteness we consider the four-dimensional case with $\Lambda=0$, an identical construction applies for Myers-Perry metrics, or for their $\Lambda$-equivalents.

Thus, in spacetime dimension four,
 let $(\mcM,\fourg)$ be the Schwarzschild metric with non-zero mass parameter.  We consider a null hypersurface $\mcN $ in  $(\mcM,\fourg)$ with two disjoint cross-sections $\secN_1$ and $\secN_2$, say $\secN_2\subset J^+(\secN_1)$. On $\secN_1$   we are given spacelike vacuum data $\CSdata{\secN_1,k}$ distinct from but close to  the data induced by $\fourg$. On $\secN_2$ we consider the family of data $\CSdata{\secN_2,k}$ arising by restriction from all
Kerr metrics.
 The goal is to find  null hypersurface data which interpolate  between $\CSdata{\secN_1,k}$ and a sufficiently small perturbation of one of the $\CSdata{\secN_2,k}$'s
  in a way such that we can carry out the spacetime gluing of Theorem~\ref{T18XII22.1}.
   The result will be  a spacetime metric which coincides with a Kerr metric in the right-wedge $\mcD(\thyp 2)$ of Figure~\ref{F17XII22.2c}. The difficulty is to arrange smallness of the perturbation of a large number of transverse derivatives of the metric at $\secN_2$.

Now, there exists $k_0<\infty$ such that   characteristic data $\CSdata{\mcN,k_0}$ which are $\epsilon$-close to the data induced by $\fourg$ will lead, through the construction of Proposition~\ref{P26XII22.1}, to a metric $\fourg_1$ which is $\epsilon$-close in $C^5$-topology to the metric $\fourg$.
The dimension-dependent number $k_0$ can be determined in principle by chasing losses of  differentiability through all the steps of the construction.
Here one uses  straightforward estimates for a hierarchical system of ODEs, where at each step a linear ODE is solved for a new field in terms of the already-determined ones.

So let $\epsilon$ be a measure of the deviation of the data  $\CSdata{\secN_1,k_0}$ from those induced by $\fourg$.

A brute-force gluing proceeds as follows: We choose a member of the Kerr family  such that $\CSdata{\secN_2,k_0}$ is $\epsilon$-close to the data induced by $\fourg$, and has the same linearly conserved radial charges as $\CSdata{\secN_1,k_0}$. We find any smoothly interpolating free data on  $J^+(\secN_1)\cap J^-(\secN_2) \subset \mcN$
which deviate from the Schwarzschild data by
$O(\epsilon)$. We use these data  in the source terms of the transport and algebraic equations of Section~\ref{s22XII22.2}, including their transverse derivatives, to obtain a  solution of these equations on $J^+(\secN_1)\cap J^-(\secN_2) $
which matches $\CSdata{\secN_2,k_0}$ up to error terms of order $\epsilon$.  If $\epsilon$ is sufficiently small,  Theorem~\ref{T18XII22.1} applies.

The spacelike-gluing version of the more sophisticated scheme of \cite{ACR1}, which appears to be critical for some applications such as \cite{CzimekRodnianski}, proceeds as follows.
The above argument works  in all dimensions, but what follows rests on work which assumes four dimensions; the higher dimensional case will be addressed elsewhere~\cite{ChCong2}.

It has been shown in~\cite{ChCong1}
how to find linearised Bondi free data  $ h_{AB}$  so that the metric $\gamma_{AB}+h_{AB}$ interpolates between  $\CSdata{\secN_1,k_0}$ and one of the data sets $\CSdata{\secN_2,k_0}$ at a linearised level.
One can then use these data  in the source terms of the transport and algebraic equations of Section~\ref{s22XII22.2} to obtain
    hypersurface data on $J^+(\secN_1)\cap J^-(\secN_2)$
     which  match $\CSdata{\secN_2,k_0}$ to order $\epsilon^2$. The gluing then follows again from Theorem~\ref{T18XII22.1}. We note that the improvement from $\epsilon$ to $\epsilon^2$ is critical for some applications, such as \cite{CzimekRodnianski},

The same arguments apply to metrics which are near to a four-dimensional Birmingham-Kottler metric with higher genus at infinity and with nonzero mass, where the mass parameter has to be adjusted to do the gluing, see~\cite[Table~1.1]{ChCong1}.

We believe that the same scheme can be used to interpolate between data with $\Lambda\in \R$ near Birmingham-Kottler data in any spacetime dimensions $n+1\ge 4$, we plan to return to this in the near future. In spacetime dimension four with $\Lambda\ne0$ the linearised analysis of~\cite{ChCong1} applies  and, in the spherical case, the same argument will lead to the desired conclusion after checking that the Kerr-(A)dS metrics provide the required family of compensating metrics.
%
%

\appendix

\section{ACR sphere data}
 \label{app22XII22.1}

Here we calculate the sphere data of \cite{ACR1} of order two in terms of Bondi section data. Given a
cross-section $\secN$ of $\mathcal{N}$,
by which we mean a submanifold of $\mathcal{N}$ intersecting all the generators of  $\mathcal{N}$ transversally,
the field
$\partial_r\equiv L$ is the field of  null  normals both
 to $\secN$ and $\mathcal{N}$, while $\underline{L}$ is the field of null normals to $\secN$ transverse to $\mathcal{N}$. In typical applications both $L$ and $\underline L$ are chosen to be future-directed, but the choice is irrelevant for the problem at hand.

 In \cite{ACR1} the  cross-section $\secN$ is chosen to be a sphere and the space-time dimension is four: both assumptions are essential for the analysis there, compare~\cite{ChCong1}.

 In Bondi coordinates we can choose 
\begin{equation}
    \label{23III22w1}
    L = \partial_r \,,\quad \underline{L} = \partial_u + U^A \partial_A -\frac{V}{2r} \partial_r
     \,.
\end{equation}
This gives
\begin{equation}
\label{23III22w2}
    \Omega := \sqrt{-\tfrac{1}{2} g(L,\underline{L})} = \frac{e^{\beta}}{\sqrt{2}}\,.
\end{equation}
The sphere data of \cite{ACR1} further involve the fields
$$
 \mbox{$\hat{L}=L/\Omega$ and $\underline{\hat{L}}=\underline{L}/\Omega$.}
$$
First, the \textit{Ricci coefficients} are defined as, for $X$ and $Y$ $\secN$-tangent vector fields,
\begin{align}
\label{23III22w3}
    \chi(X,Y) & = g(\nabla_X \hat L, \, Y)  \,, &  \underline{\chi} &= g(\nabla_X \hat{\underline{L}},\, Y)\,, \nonumber
\\
    \zeta(X) &= \frac{1}{2} g(\nabla_X \hat{L},\,\hat{\underline{L}})\,, & \underline{\zeta}(X) &= \frac{1}{2} g(\nabla_X \hat{\underline{L}},\,\hat{L}) \,,\nonumber
\\
    \eta &= \zeta + d \log \Omega\,, & \underline{\eta} &= -\zeta + d \log \Omega\,,\nonumber
\\
    \omega &= D \log \Omega\,, &  \underline{\omega} &= \underline{D} \log \Omega \,,
\end{align}
where $D$ and $\underline{D}$ respectively denote the projection of the Lie derivative along $L$ and $\underline{L}$ onto the tangent space of $\secN$. The \textit{null curvature components} involved in the sphere data are
\begin{equation}
\label{23III22w4}
    \alpha(X,Y) = R(X,\hat{L},Y,\hat{L})\,,\quad \underline{\alpha}(X,Y) = R(X,\hat{\underline{L}},Y,\hat{\underline{L}}).
\end{equation}
The $C^2$-sphere data of \cite{ACR1} is the collection of fields
\begin{equation}
    \label{23III22w5}
    (\Omega,\, \cancel{g}, \, \Omega \tr \chi,\, \hat\chi,\, \Omega \tr\underline\chi\,,\hat{\underline{\chi}}\,,\eta\,,\omega,\, D\omega,\, \underline\omega,\, \underline{D} \underline{\omega},\, \alpha, \underline{\alpha})\,,
\end{equation}
where $\tr$ denotes the trace with respect to the metric $\cancel{g}$ on $\secN$ and the hat $\hat{\cdot}$ above a tensor denotes the traceless part.
Leting $\snD$ denote the covariant derivative of the metric $\cancel{g}$,
In Bondi coordinates the fields \eqref{23III22w5} read
\begin{align}
    \label{23III22w6}
    \Omega
    =&
     \frac{e^{\beta}}{\sqrt{2}}\,,\quad  \cancel{g} = r^2\gamma
     \,,
      \nonumber
\\
    \chi_{AB}
    = & \frac{1}{\Omega}
      ( r
       \gamma_{AB}  + \tfrac{1}{2} r^2 \partial_r\gamma_{AB})
        \equiv  \frac{1}{2 \Omega  }
      \partial_r( r^2 \gamma_{AB})
       \,,
       \nonumber
\\
       \underline{\chi}_{AB}
       = & - \frac{2 \gamma_{AB} V -  4 \snD_{(A}U_{B)}+ r ( V \partial_r\gamma_{AB} - 2 r \partial_u\gamma_{AB})}{2 \sqrt{2} e^{\beta}}\,,
\\
    \eta_A
    = &
    \partial_{A}\beta -  \frac{\gamma_{AB} r^2 \partial_{{r}{}}U^{B}}{2 e^{2 \beta}}\,, \quad \omega = \partial_r \beta\,, \quad D\omega  = \partial^2_r\beta \,,
   \quad
    \underline{\omega} = U^{A} \partial_{A}\beta -  \frac{V \partial_{{r}{}}\beta}{2 r} + \partial_{{u}{}}\beta \,,
\\
    \underline{D}\underline{\omega}
    = &
     - \frac{U^{A} V \partial_{A}\partial_r\beta}{r} + 2 U^{A} \partial_{A}\partial_u\beta + U^{A} \partial_{A}U^{B} \partial_{B}\beta + U^{A} U^{B} \partial_{B}\partial_{A}\beta -  \frac{V^2 \partial_r\beta}{4 r^3} -  \frac{U^{A} \partial_{A}V \partial_r\beta}{2 r}
      \nonumber
\\
    &-  \frac{V \partial_{A}\beta \partial_rU^{A}}{2 r} + \frac{V \partial_r\beta \partial_rV}{4 r^2} + \frac{V^2 \partial_r^2\beta}{4 r^2} -  \frac{V \partial_r\partial_u\beta}{r} + \partial_{A}\beta \partial_uU^{A} -  \frac{\partial_r\beta \partial_uV}{2 r} + \partial_u^2\beta \,,
\\
    \alpha_{AC}   = &
    \frac{4 \gamma_{AC} r \partial_r\beta}{e^{2 \beta}} -  \frac{2 r \partial_r\gamma_{AC}}{e^{2 \beta}} + \frac{2 r^2 \partial_r\beta \partial_r\gamma_{AC}}{e^{2 \beta}} + \frac{\gamma^{BD} r^2 \partial_r\gamma_{AB} \partial_r\gamma_{CD}}{2 e^{2 \beta}} -  \frac{r^2 \partial_r^2\gamma_{AC}}{e^{2 \beta}}
    \,,
\\
    \underline{\alpha}_{AC}
      = &   e^{-2\beta}\bigg( r^{-1} V \snD_{(A} \partial_r U_{B)} -  2 \snD_{(A}\partial_u U_{B)} + \frac{e^{2\beta}\snD_{B}\snD_{A}V}{r}  -  2 U^{C} \snD_{C}\snD_{(A}U_{B)} \nonumber \\
&  -  \frac{V^2 \partial_r^2\gamma_{AB}}{4 } + r V \partial_r\partial_u\gamma_{AB}\nonumber  -  r^2 \partial_u^2\gamma_{AB}\bigg)
      \nonumber
\\
        &
       + \tilde{\underline{\alpha}}_{AC}[r, r^{-1}, \gamma_{BD}, \gamma^{BD}, \partial \gamma_{BD},
       e^{-2\beta}, \partial \beta,  U_B, \partial U_B,    V, \partial V, \partial_u \gamma_{BD},  \partial_u \beta]
      \,,
\end{align}
with a polynomial function $\tilde{\underline{\alpha}}_{AC}$ of the arguments indicated; the explicit formula is not very enlightening and too long to be usefully displayed.
We use ``$\partial $'' in the arguments of $\tilde{\underline{\alpha}}_{AC}$ to denote  $\partial_r$ and $\partial_A$ derivatives, with $\partial_u$ derivatives indicated explicitly there. 

\section{Bondi coordinates anchored at $\secN$}
 \label{s18IV22.1asdf}

The construction of Bondi coordinates in four spacetime dimensions starting from $\scri$ is well known~\cite{GerochWinicour81}, and generalises immediately to higher dimensions. We indicate here how to adapt  the construction to our setting, to make clear the freedom involved.

Let $\secN$ be a cross-section of a smooth, null, connected hypersurface $\mcN$ in an $(n+1)$-dimensional  spacetime $(\mcM,g)$.
Let $\umcN$ be a null hypersurface such that $\mcN\cap \umcN = \secN$, with transverse intersection.
(Thus $\secN$ is spacelike, with both $T\mcN $ and $T\umcN$ orthogonal to $T\secN$.)
Let $\tilde x^A$ denote local coordinates on $\secN$.
We consider, first, Isenberg-Moncrief~\cite{VinceJimcompactCauchyCMP} coordinates $(\tilde u,\btdv,\tilde x^C)$ around
$$
 \secN = \{ \tilde u=0=\btdv\}
 \,,
$$
and with the metric taking the form
 \ptcheck{3I23; crosschecked with Proposition 4.3.14 of \cite{ChBlackHoles}}
\begin{equation}\label{18IV22.1}
  \fourg =
  \big(
   \btdv   \, \red{\alpha} \,\red{\mathrm{d}}\tilde  u - 2 \red{\mathrm{d}}\btdv
  + 2  \,\btdv   \, \red{\beta}_A \red{\mathrm{d}}\tilde  x^A
  \big)
 \red{\mathrm{d}}\tilde  u + g_{AB}\red{\mathrm{d}}\tilde  x^A \red{\mathrm{d}}\tilde  x^B
  \,,
\end{equation}
for some fields $\red{\alpha}$ and $\red{\beta}_A$. Here we have denoted by $(\tilde u, \btdv, \tilde x^A)$ the coordinates $(\newu, \newr, x^A)$
of \eqref{10I22.w1}, to avoid confusion with the $(u,r,x^A)$ Bondi coordinates that we are about to construct.

Note that while the $\tilde x^A$'s are local coordinates, the coordinate functions $\tilde u$ and $\btdv$ are defined globally in a neighborhood of $\secN$. The level sets of the coordinate $\tilde u$ are null hypersurfaces and we denote $\mcN:=\{\tilde u=0\}$ in this Appendix;%
\footnote{We caution that this differs from the notation in Section \ref{s6I23.1}, where $\mcN$ was chosen to be $\{\btdv=0\}$.}
we will be constructing Bondi coordinates $(u,r,x^A)$ such that $\mcN = \{u=0\}$.
The hypersurface $\umcN:=\{\btdv=0\}$ is also null, but not necessarily so the hypersurfaces $\btdv\ne 0$.
The sign of $\fourg _{\tilde u \btdv}$ has been determined by our signature $(-,+,\ldots,+)$ together with the requirement that $\partial_{\tilde u}$ and $\partial_\btdv$ are consistently time-oriented at $\secN$, say future oriented.

In the  Isenberg-Moncrief construction one can take
  the integral curves of $\ell:= \partial_\btdv |_\mcN$  to be affinely-parameterised future-directed null geodesics
  (in which case $\red{\alpha}$ vanishes on $\mcN$),
  then the coordinate system above is uniquely defined up to the choice of this last parameterisation. In order to get rid of this freedom, consider the divergence $\tthetatau$   of $\mcN$  defined in \eqref{dfn_tau},
where we decorate $\thetatau$ with a tilde to emphasise its dependence upon the coordinate $\btdv$.
Under the rescaling $\ell \mapsto f(\tilde  x^A) \ell$   we have
\begin{equation}\label{4I23.1}
     \tthetatau  =\frac 12  g^{AB}\partial_{\btdv} g_{AB}
      \mapsto \frac f 2  g^{AB}\partial_{\btdv} g_{AB} = f \tthetatau
  \,.
\end{equation}
Assuming that \emph{$\tthetatau$ has no zeros on $\secN$}, we can choose a unique function $f>0$ so that, after the above rescaling has been done, the new function $\tthetatau$ satisfies

\begin{equation}\label{4I23.2}
     \tthetatau|_\secN  = \pm (n-1)
  \,,
\end{equation}
thus preserving the future-directed character of $\ell$, or choose a unique $f$ so that
\begin{equation}\label{4I23.3}
     \tthetatau|_\secN  =  (n-1)
\end{equation}
if the time-orientation of $\ell$ is ignored.
%
%

The field
$$
 \momega:=\sqrt {\det g_{AB}} |_{\secN}
$$
defines a scalar density  on $\secN$. We  extend $\momega$ to $\umcN$ by requiring $\mcL_{\partial_{\tilde u}} \momega|_{\umcN} = 0$, and then we  extend it away from $\umcN$ by  requiring $\mcL_{\partial_{\btdv}} \momega = 0$. Still denoting by $\mu$ the field so extended, since $\partial_\btdv$ and $\partial_{\tilde u}$ commute we find that $
 \mcL_{\partial_{\btdv}}  \mcL_{\partial_{\tilde u}} \momega = 0
$, which further implies
$$
 \mcL_{\partial_{\btdv}}  \momega = \mcL_{\partial_{\tilde u}} \momega = 0
$$
throughout the domain of definition of the coordinates.

We define a function $r=r(\tilde u,\btdv,\tilde x^A)$ by the formula
\begin{equation}\label{18IV22.2as}
  r ^{1-n}  :=  \frac{\momega}{\sqrt {\det g_{AB}} }
   \,.
\end{equation}
Note that
\begin{equation}\label{4I23.1b} r |_{\secN} =1
 \,.
\end{equation}

We wish to replace the coordinates $(\tilde u,\btdv,\tilde x^A)$ by
$$
 (u=\tilde u,r,x^A = \tilde x^A)
  \,.
$$
Using monotonicity and the implicit function theorem, we see that this is possible on the  set where
\begin{equation}\label{3I23.1}
  0\ne  \partial_{\btdv} r
   =  \frac{r}{2(n-1)}    g^{AB}\partial_{\btdv} g_{AB}
  \,.
\end{equation}
This is directly related to the divergence $ \tthetatau$ of $\mcN$:
\begin{equation}\label{3I23.1d}
    \partial_{\btdv} r
   \Big|_\mcN =   \frac{r  \tthetatau}{ (n-1)}
  \,.
\end{equation}
We thus obtain a well behaved coordinate system  $(r,x^A)$ on  $\mcN$,   and $(u,r,x^A)$ near $\mcN$, unless  $r$ becomes zero, which happens e.g.\ at the vertex of a light cone, or unless $\tthetatau$ acquires a zero. Hence we restrict ourselves to the subset of $\mcN$ where  $r>0$ and $|\tthetatau|>0$.

Similarly to \eqref{3I23.1} we have
\begin{equation}\label{3I23.1c}
  \frac{\partial r}{\partial \tilde u } =  \frac{r}{2(n-1)}    g^{AB}\partial_{\tilde u} g_{AB}  \equiv   \frac{r\tuthetatau}{ (n-1)}
  \,,
\end{equation}
where $ \tuthetatau$ is the expansion of the level sets of $\btdv$.

We note that the vector field $\partial_{\tilde u}$ is uniquely determined on $\secN$ by the requirement that $\partial_{\tilde u}$ is orthogonal to $T\secN$ and satisfies $\fourg (\partial_{\tilde u},\partial_{\btdv})|_\secN=-1$. Hence $\tuthetatau|_{\secN}$ is uniquely determined by the requirement \eqref{4I23.2} and by $\secN$,
without  the need to introduce the null transverse hypersurface $\umcN$.  The function $\tuthetatau|_{\secN}$ is sometimes called the \emph{null mean curvature of $\secN$ along  $\partial_{\tilde u}$}.

The change of coordinates $(\tilde u,\btdv,\tilde x^A)\mapsto (u,r,x^A)$
 brings the metric to the form
\begin{equation}\label{18IV22.1123}
  \fourg =
  \,\btdv   \, \red{\alpha} \, \mathrm{d} u^2 - 2\mathrm{d}u \, (\partial_{u} \btdv \, \mathrm{d}u +
 \partial_r \btdv \,\mathrm{d}r +
 \partial_A \btdv\, \mathrm{d}x^A)
  + 2 \,\btdv   \, \red{\beta}_A \mathrm{d}x^A
  \mathrm{d}u + g_{AB}\mathrm{d}x^A \mathrm{d}x^B
  \,,
\end{equation}
which can be rewritten using the Bondi parameterisation
\begin{eqnarray}
\fourg
 &  =  &-\frac{V}{r}e^{2\beta} \mathrm{d}u^2-2 e^{2\beta}\mathrm{d}u\,\mathrm{d}r
   +r^2\zhTBW_{AB}\Big(\mathrm{d}x^A-U^A\mathrm{d}u\Big)\Big(\mathrm{d}x^B-U^B\mathrm{d}u\Big)
    \, ,
     \label{20IV22.1}
\end{eqnarray}
where it is assumed that $\partial_r$ and $\partial_u$ are consistently
time-oriented at $\secN$.

From \eqref{4I23.3} and \eqref{3I23.1} we find
\begin{equation}\label{3I23.1b}
   \partial_{\btdv} r \Big|_\secN =  \pm 1
\end{equation}
if and only if \eqref{4I23.2} holds. Assuming the associated parameterisation of the generators of $\mcN$ we obtain
\begin{equation}\label{4I23.5a}
  \frac{\partial(r, u, x^A)}{\partial(\btdv,\tilde u,\tilde x^B)}\Big|_\secN
  = \left(
      \begin{array}{ccc}
        \pm 1 &  \partial_{\tilde u} r  & 0 \\
        0 & 1 & 0 \\
        0 & 0 & \delta^A_B \\
      \end{array}
    \right)
  \,.
\end{equation}
Hence
\begin{equation}\label{4I23.5b}
  \frac{\partial(\btdv,\tilde u,\tilde x^B)}{\partial(r, u, x^A)}\Big|_\secN
  = \left(
      \begin{array}{ccc}
        \pm 1 & \mp  \partial_{\tilde u} r  & 0 \\
        0 & 1 & 0 \\
        0 & 0 & \delta^B_A \\
      \end{array}
    \right)
  \,.
\end{equation}
This leads to the following form of~\eqref{18IV22.1123}  at $\secN$
\begin{equation}\label{4I23.6}
  \fourg|_\secN  = \mp  2\mathrm{d}u \, ( -   \partial_{\tilde u} r \, \mathrm{d}u
  + \mathrm{d}r ) + g_{AB}\mathrm{d}x^A \mathrm{d}x^B
  \,,
\end{equation}
which together with \eqref{3I23.1c}, after changing $u$ to $- u$ if necessary,
shows that  at $\secN$ it holds
\begin{equation}\label{20IV22.2}
  U_A|_{\secN} := -r^{-2}   \fourg(\partial_u,\partial_A) \big|_{\secN}
   \equiv  \zhTBW_{AB} U^B  \big|_{\secN}
   =0
    \,,
    \quad
     \beta|_{\secN} = 0
    \,,
    \quad
    V|_{\secN} =
     - \frac{2 \tuthetatau}{n-1}
  \,.
\end{equation}

Summarising, we have proved:

\begin{Proposition}
 \label{P4I23.1}
Let $\mcN$ be a null hypersurface with $\theta>0$ and suppose that $\mcN$ contains a smooth submanifold   $\secN$ which meets every generator of $\mcN$ transversally and precisely once. There exists a unique coordinate system near $\mcN$  in which the metric takes the Bondi form \eqref{20IV22.1} with $\mcN=\{u=0\}$ and in which
\begin{equation}\label{4I23.31}
  r|_\secN=1
  \,,
   \qquad \beta|_\secN = 0
  \,.
\end{equation}
In this coordinate system we have
\begin{equation}\label{4I23.32}
    U^A|_\secN=0
    \,,
    \qquad
      \tthetatau|_\secN  =  (n-1)
    \,,
    \qquad
    V|_{\secN} =  - \frac{2 \tuthetatau}{n-1}
  \,,
\end{equation}
where $\tthetatau$ is the expansion of $\{u=0\}$ with respect to the
 affinely-normalised
geodesic null vector field $\partial_{\btdv}$, and $\tuthetatau$ is that of the level sets of $\btdv$ with respect to the Isenberg-Moncrief geodesic vector field $\partial_{\tilde u}$.
\qed
\end{Proposition}

The coordinates of Proposition~\ref{P4I23.1} will be referred to as \emph{Bondi coordinates adapted to $\mcN$ anchored at $\secN$.}

\begin{Remark}
 \label{R5I23.1}
{\rm
The proof of Proposition~\ref{P4I23.1} applies word-for-word in spacetimes in which $\mcN$ is a smooth boundary, or in spacetimes with a boundary consisting of two null hypersurfaces $\mcNone$ and $\mcNtwo$ intersecting at $\secN$, in which cases   the coordinates are only defined  on one side of $\mcN$.
}
\qed
\end{Remark}

{
\begin{remark}
 \label{R25I23}
{\rm
The Isenberg-Moncrief construction can be carried-out using any parameterisation of the generators of $\mcN$, not necessarily affine.
Likewise our construction above applies for any parameterisation of the generators, which shows that there is a lot of residual coordinate freedom in the Bondi form of the metric near a null hypersurface. This explains in particular why the Bondi hypersurface data of
Section~\ref{s22XII22.2} have more freedom than the Isenberg-Moncrief hypersurface data of Section~\ref{s6I23.1}.
}
\qed
\end{remark}
}

More generally, whether or not \eqref{4I23.2} holds we have
\begin{equation}\label{4I23.5c}
  \frac{\partial(r, u, x^A)}{\partial(\btdv,\tilde u,\tilde x^B)}
  = \left(
      \begin{array}{ccc}
         \partial_{\btdv} r&  \partial_{\tilde u} r  & \frac{\partial r}{\partial \tilde x^B}\\
        0 & 1 & 0 \\
        0 & 0 & \delta^A_B \\
      \end{array}
    \right)
  \,,
\end{equation}
so that
\begin{equation}\label{4I23.5d}
  \frac{\partial(\btdv,\tilde u,\tilde x^B)}{\partial(r, u, x^A)}
  = \left(
      \begin{array}{ccc}
        \big( \partial_{\btdv} r
        \big)^{-1}
         \ & \
        -\big( \partial_{\btdv} r
        \big)^{-1} \partial_{\tilde u} r
         \ & \ \
        -\big( \partial_{\btdv} r
        \big)^{-1} \partial_{\tilde x^A} r
         \\
        0 & 1 & 0 \\
        0 & 0 & \delta^B_A \\
      \end{array}
    \right)
  \,,
\end{equation}
and~\eqref{18IV22.1123}  reads
\begin{eqnarray}
  \fourg
   & = &
 \,\btdv   \, \red{\alpha} \, \mathrm{d} u^2
  - \frac{2
        \Big[ \mathrm{d}r
        -
  \partial_{\tilde u} r\,\mathrm{d} u
 -  \partial_{\tilde x^A} r\, \mathrm{d}x^A
 \Big]  }{ \partial_{\btdv} r}
         \mathrm{d}u
  + 2 \,\btdv   \, \red{\beta}_A \mathrm{d}x^A
  \mathrm{d}u + g_{AB}\mathrm{d}x^A \mathrm{d}x^B
  \nonumber
\\
   & = &
 \Big(
  \, \btdv \, \red{\alpha} + 2\frac{
  \partial_{\tilde u} r }{ \partial_{\btdv} r}
  \Big) \mathrm{d} u^2
  - \frac{2
          \mathrm{d}r\, \mathrm{d}u  }{ \partial_{\btdv} r}
  + 2
   \Big( \btdv   \, \red{\beta}_A  +    \frac{  \partial_{\tilde x^A} r  }{ \partial_{\btdv} r}
   \Big)
         \mathrm{d}u\, \mathrm{d}x^A
   + g_{AB}\mathrm{d}x^A \mathrm{d}x^B
  \,.
 \label{8I23.2}
\end{eqnarray}
This, together with \eqref{3I23.1d}-\eqref{3I23.1c}, after changing $u$ to $- u$ if necessary,
shows that
\begin{align}\label{8I23.3}
     \beta|_{\mcN} &=
     \frac{1}{2} \ln \frac{n-1}{r \tthetatau}
    \,,
    \quad
  U_A|_{\mcN} =
  -\frac{\btdv}{r^2} {\beta}_A
  -  \frac{(n-1)}{  r^3 \tthetatau}
   \partial_{\tilde x^A} r
    \,,
    \nonumber
\\
    V|_{\mcN} &=
     -\frac{2 r^2 \tuthetatau}{n-1}
      -\frac{\btdv  {\alpha} r^2 \tthetatau}{n-1}
     + \frac{r^2 \btdv^2 g^{AB}\red{\beta}_A \red{\beta}_B \, \tthetatau}{n-1}
     +\frac{n-1}{\tthetatau} g^{AB} \partial_{\tilde x^A} r \,\partial_{\tilde x^B} r
  \,.
\end{align}

In particular we see that   $r=\xtwo$ in Theorem~\ref{T5I23.1} is  possible if and only if 
\begin{equation}\label{8I23.6}
   \partial_{\tilde x^A} r \big|_{\secN}
   =0
    \,,
    \quad
  \tuthetatau\big|_{\secN} = 0
  \,.
\end{equation}

\section{A variational identity}
 \label{CWs23II22.1}

The aim of this Appendix is to show that the restriction of spacetime Killing vectors to a hypersurface lies in the kernel of the adjoint of the linearisation of the vacuum constraints operator. This shows in particular that spacetime Killing vectors provide obstructions to characteristic gluing  based on the implicit function theorem.

Let $\fourg $ be a solution of the vacuum Einstein equations  with a cosmological constant $\red{\Lambda}\in \R$. Let $\hyp$ be a hypersurface of any causal type, possibly with boundary, and let 
 $\lambda \mapsto \fourg(\lambda)$ be a family of Lorentzian metrics along $\hyp$ depending differentiably on a parameter $\lambda$ such that  $\fourg \equiv \fourg(0)$.  The variational operator $\delta$ is defined by evaluating $\frac \partial{\partial\lambda}$ at $\lambda=0$.
We set
\begin{eqnarray}
  &
 \mcE^{\red{\alpha\beta}}:=
    -\frac{\sqrt{|\fourg|}}{16 \pi}\left({G}^{\red{\alpha\beta}} + \red{\Lambda} \fourg^{\red{\alpha\beta}}\right)
     \,,
 &
\\
 &
 {\pi}^{\red{\alpha\beta}} := \frac 1{16 \pi} \sqrt{|\fourg|} \  \fourg^{\red{\alpha\beta}}
  \,,
 &
\\
 &
A^{\lambda}_{\red{\alpha\beta}} := {\Gamma}^{\lambda}_{\red{\alpha\beta}} -
{\delta}^{\lambda}_{(\alpha} {\Gamma}^{\kappa}_{\beta ) \kappa}
 \, ,
 &
\end{eqnarray}
where $G_{\alpha\beta}$ is the Einstein tensor and $\red{\Lambda} $ the cosmological constant.
The following variational identity has been proved in~\cite{CJKKerrdS},%
\footnote{The reader might notice that~\cite{CJKKerrdS} uses a non-standard convention on the sign of the cosmological constant, opposite to the one here.}
see Equation~(2.27) there:
\begin{eqnarray}
 \nonumber
 \lefteqn{ ({\mycal L}_X A^{\lambda}_{\red{\alpha\beta}}) \delta {\pi}^{\red{\alpha\beta}} - ({\mycal L}_X{\pi}^{\red{\alpha\beta}}) \delta  A^{\lambda}_{\red{\alpha\beta}}
   =  -2  X^\red{\mu} \delta {\cal E}^\lambda{}_{\red{\mu}}
   + X^\lambda {\cal E}^ {\red{\alpha\beta}}  \red{\delta \fourg}_{\red{\alpha\beta}}
   }
   &&
\\
 \nonumber
 &&
   +\frac {1}{16 \pi} \partial_\mu \bigg\{
 \delta\big[ \sqrt{|\fourg|} (\nabla^\mu X^\lambda - \nabla^\lambda
 X^\mu ) \big]
 - \sqrt{|\fourg|} (\nabla^\mu \delta X^\lambda - \nabla^\lambda
 \delta X^\mu )
 \bigg\}
\\
 & &
+ \partial_\red{\mu} \left\{ X^\lambda {\pi}^{\red{\alpha\beta}} \delta A^{\red{\mu}}_{\red{\alpha\beta}}
 -X^\red{\mu} {\pi}^{\red{\alpha\beta}} \delta A^{\lambda}_{\red{\alpha\beta}} \right\}
\, .
   \label{CWdL-Waldbis}
\end{eqnarray}
Integrating \eqref{CWdL-Waldbis} over $\hyp$, assuming that the integral converges, and  that  the metrics $\fourg (\lambda)$ coincide with $\fourg  $   near the boundary of $\partial\hyp$,  if any, one
finds
\begin{eqnarray}
  \int_{\hyp}
  \big(
  ({\mycal L}_X A^{\lambda}_{\red{\alpha\beta}}) \delta {\pi}^{\red{\alpha\beta}} - ({\mycal L}_X{\pi}^{\red{\alpha\beta}}) \delta  A^{\lambda}_{\red{\alpha\beta}}
  \big) dS_\lambda
   =  -2   \int_{\hyp}
   X^\red{\mu} \delta {\cal E}^\lambda{}_{\red{\mu}}dS_\lambda
\, .
   \label{CWdL-Waldbisa}
\end{eqnarray}

Let $N_\lambda$ be any field of conormals to $\hyp$, thus $\delta {\cal E}^\lambda{}_{\red{\mu}}$ enters this identity only through the components $\delta {\cal E}^\lambda{}_{\red{\mu}}N_\lambda$. The equations $ {\cal E}^\lambda{}_{\red{\mu}}N_\lambda=0$ are the constraint equations on $\hyp$, and \eqref{CWdL-Waldbisa} expresses the well-known fact that, for spacelike $\hyp$'s, the constraint equations provide an action principle for the Einstein equations. This remains true for characteristic hypersurfaces in view of \eqref{CWdL-Waldbisa}, but is perhaps somewhat less known; compare~\cite{KorbiczTafel,CJKbhthermo}.

The operator $\delta {\cal E}^\lambda{}_{\red{\mu}}N_\lambda$ is the linearisation of the constraint equations on $\hyp$ acting on linearised gravitational initial data on $\hyp$.  Integration by parts reexpresses the right-hand side as the adjoint operator of the linearised constraint equations acting on $X$.

Now, Killing vectors in space-time annihilite the left-hand side of \eqref{CWdL-Waldbisa}. It follows that   Killing vectors of the spacetime metric are in the kernel of this operator, in the following sense:
\emph{if $X$ is a vector field satisfying the Killing equations and their first derivatives on  $\mcN$}, then
\begin{eqnarray}
   \int_{\hyp}
   X^\red{\mu} \delta {\cal E}^\lambda{}_{\red{\mu}}dS_\lambda
   =0
   \label{CWdL-Waldbisb}
\end{eqnarray}
for all variations $\red{\delta \fourg}_{\mu\nu}$ as described.

It is known, for spacelike $\hyp$'s,  that spacetime Killing vectors exhaust  the kernel~\cite{Moncrief75}: the left-hand side of \eqref{CWdL-Waldbisb} vanishes for all variations as above
\emph{if and only if $X$ is a vector field satisfying the Killing equations and their first derivatives on  $\mcN$}.
Our gluing results in this paper suggest strongly that this remains true for characteristic hypersurfaces, but this remains to be seen; compare~\cite{ChPaetzKIDs}.

\bibliographystyle{amsplain}

\bibliography{ChCong}

\providecommand{\bysame}{\leavevmode\hbox to3em{\hrulefill}\thinspace}
\providecommand{\MR}{\relax\ifhmode\unskip\space\fi MR }
\providecommand{\MRhref}[2]{%
  \href{http://www.ams.org/mathscinet-getitem?mr=#1}{#2}
}
\providecommand{\href}[2]{#2}
\begin{thebibliography}{10}

\bibitem{ACR3}
S.~Aretakis, S.~Czimek, and I.~Rodnianski, \emph{{Characteristic gluing to the
  Kerr family and application to spacelike gluing}},  (2021), arXiv:2107.02456
  [gr-qc].

\bibitem{ACR1}
\bysame, \emph{{The characteristic gluing problem for the Einstein equations
  and applications}},  (2021), arXiv:2107.02441 [gr-qc].

\bibitem{ACR2}
\bysame, \emph{{The characteristic gluing problem for the Einstein vacuum
  equations. Linear and non-linear analysis}},  (2021), arXiv:2107.02449
  [gr-qc].

\bibitem{BarrabesIsrael}
C.~Barrab\`es and W.\ Israel, \emph{Thin shells in general relativity and
  cosmology: {T}he lightlike limit}, Phys.\ Rev.\ D \textbf{43} (1991),
  1129--1142.

\bibitem{Bartnik:quasi-sph}
R.~Bartnik, \emph{Quasi-spherical metrics and prescribed scalar curvature},
  Jour.\ Diff.\ Geom. \textbf{37} (1993), 31--71. \MR{93i:53041}

\bibitem{BBM}
H.~Bondi, M.G.J. van~der Burg, and A.W.K. Metzner, \emph{Gravitational waves in
  general relativity {VII}: Waves from axi--symmetric isolated systems}, Proc.\
  Roy.\ Soc.\ London A \textbf{269} (1962), 21--52. \MR{MR0147276 (26 \#4793)}

\bibitem{Borel2}
E.~Borel, \emph{Addition au m\'emoire sur les s\'eries divergentes}, Ann. Sci.
  \'Ecole Norm. Sup. (3) \textbf{16} (1899), 132--136. \MR{MR1508966}

\bibitem{BorelMem}
E.~Borel, \emph{M\'{e}moire sur les s\'{e}ries divergentes}, Ann. Sci.
  \'{E}cole Norm. Sup. (3) \textbf{16} (1899), 9--131. \MR{1508965}

\bibitem{CCW}
A.~Cabet, P.T. Chru\'{s}ciel, and R.~Tagne~Wafo, \emph{On the characteristic
  initial value problem for nonlinear symmetric hyperbolic systems, including
  {E}instein equations}, Dissertationes Math.\ (Rozprawy Mat.) \textbf{515}
  (2016), 72 pp., arXiv:1406.3009 [gr-qc]. \MR{3528223}

\bibitem{CederbaumExtensions}
A.J. Cabrera~Pacheco, C.~Cederbaum, S.~McCormick, and P.~Miao,
  \emph{Asymptotically flat extensions of {CMC} {B}artnik data}, Class.\
  Quantum Grav. \textbf{34} (2017), 105001, 15. \MR{3649699}

\bibitem{CarlottoLR}
A.~Carlotto, \emph{{The general relativistic constraint equations}}, Living
  Rev. Rel. \textbf{24} (2021), no.~1, 2.

\bibitem{CCM2}
Y.~Choquet-Bruhat, P.T. Chru\'{s}ciel, and J.M. Mart\'in-Garc\'ia, \emph{{The
  Cauchy problem on a characteristic cone for the Einstein equations in
  arbitrary dimensions}}, Ann.\ H.\ Poincar\'e \textbf{12} (2011), 419--482,
  arXiv:1006.4467 [gr-qc]. \MR{2785136}

\bibitem{ChruscielSigma}
P.~T. Chru\'{s}ciel, \emph{{The existence theorem for the general relativistic
  Cauchy problem on the light-cone}}, Forum Math. Sigma \textbf{2} (2014),
  Paper No. e10, 50. \MR{3264243}

\bibitem{ChBourbaki}
P.T. Chru\'{s}ciel, \emph{{Anti-gravity \`a la Carlotto-Schoen}},
  Ast\'{e}risque \textbf{1120} (2019), no.~407, Exp. No. 1120, 1--25,
  S\'{e}minaire Bourbaki. Vol. 2016/2017. Expos\'{e}s 1120--1135,
  arXiv:1611.01808 [math.DG]. \MR{3939271}

\bibitem{ChCong1}
P.T. Chru\'{s}ciel and W.~Cong, \emph{Characteristic gluing with {$ \Lambda $:
  1. Linearised Einstein} equations on four-dimensional spacetimes},  (2022),
  arXiv:2212.10052 [gr-qc].

\bibitem{ChCong2}
P.T. Chru\'{s}ciel, W.~Cong, and F.~Gray, \emph{Characteristic gluing with {$
  \Lambda $ 2. Linearised Einstein} equations in higher dimension},  (2023).

\bibitem{ChDelay}
P.T. Chru\'{s}ciel and E.~Delay, \emph{On mapping properties of the general
  relativistic constraints operator in weighted function spaces, with
  applications}, M\'em.\ Soc.\ Math.\ de France. \textbf{94} (2003), 1--103
  (English), arXiv:gr-qc/0301073. \MR{MR2031583 (2005f:83008)}

\bibitem{ChDelayAH}
\bysame, \emph{Gluing constructions for asymptotically hyperbolic manifolds
  with constant scalar curvature}, Commun.\ Anal.\ Geom. \textbf{17} (2009),
  343--381, arXiv:0711.1557[gr-qc]. \MR{2520913 (2011a:53052)}

\bibitem{CJKKerrdS}
P.T. Chru\'{s}ciel, J.~Jezierski, and J.~Kijowski, \emph{{Hamiltonian dynamics
  in the space of asymptotically Kerr-de Sitter spacetimes}}, Phys.\ Rev.
  \textbf{D92} (2015), 084030, 30 pp., arXiv:1507.03868 [gr-qc]. \MR{3459465}

\bibitem{ChPaetz}
P.T. Chru\'{s}ciel and T.-T. Paetz, \emph{{The many ways of the characteristic
  Cauchy problem}}, Class.\ Quantum Grav. \textbf{29} (2012), 145006, 27 pp.,
  arXiv:1203.4534 [gr-qc]. \MR{2949552}

\bibitem{ChPaetzKIDs}
P.T. Chru\'{s}ciel and T.-T. Paetz, \emph{{KIDs like cones}}, Class.\ Quantum
  Grav. \textbf{30} (2013), 235036, arXiv:1305.7468 [gr-qc].

\bibitem{ChruscielWafoGray}
P.T. Chru\'{s}ciel, R.~Tagne~Wafo, and F.~Gray, \emph{{The ''neighborhood
  theorem'' for the general relativistic characteristic Cauchy problem in
  higher dimension}},  (2023), arXiv:2305.07306 [gr-qc].

\bibitem{Collingbourne}
S.~Collingbourne, \emph{The {Gregory-Laflamme} instability and conservation
  laws for linearised gravity}, Ph.D. thesis, University of Cambridge, 2022,
  \url{https://api.repository.cam.ac.uk/server/api/core/bitstreams/6a5215f2-5719-4ca4-8f28-7a4131482097/content}.

\bibitem{CortierKdS}
J.~Cortier, \emph{{Gluing construction of initial data with Kerr-de Sitter
  ends}}, Ann.\ H.\ Poincar\'e \textbf{14} (2013), 1109--1134, arXiv:1202.3688
  [gr-qc]. \MR{3070748}

\bibitem{Corvino}
J.~Corvino, \emph{Scalar curvature deformation and a gluing construction for
  the {E}instein constraint equations}, Commun.\ Math.\ Phys. \textbf{214}
  (2000), 137--189. \MR{MR1794269 (2002b:53050)}

\bibitem{CorvinoSchoen2}
J.~Corvino and R.~Schoen, \emph{On the asymptotics for the vacuum {E}instein
  constraint equations}, Jour.\ Diff.\ Geom. \textbf{73} (2006), 185--217,
  arXiv:gr-qc/0301071. \MR{MR2225517 (2007e:58044)}

\bibitem{Czimek:2016ydb}
S.~Czimek, \emph{{An extension procedure for the constraint equations}}, Ann.
  PDE \textbf{4} (2018), no.~1, Paper No. 2, 130. \MR{3740633}

\bibitem{CzimekRodnianski}
S.~Czimek and I.~Rodnianski, \emph{{Obstruction-free gluing for the Einstein
  equations}},  (2022), arXiv: 2210.09663 [gr-qc].

\bibitem{CJKbhthermo}
E.~Czuchry, J.~Jezierski, and J.~Kijowski, \emph{Dynamics of a gravitational
  field within a wave front and thermodynamics of black holes}, Phys.\ Rev. D
  \textbf{70} (2004), 124010, 14, arXiv:gr-qc/0412042. \MR{2124700
  (2005k:83071)}

\bibitem{Frittelli:2004pk}
S.~Frittelli, \emph{{Well-posed first-order reduction of the characteristic
  problem of the linearized Einstein equations}}, Phys.\ Rev.\ D \textbf{71}
  (2005), 024021, 7, arXiv:gr-qc/0408035. \MR{2125656}

\bibitem{Frittelli:1999yr}
S.~Frittelli and L.~Lehner, \emph{{Existence and uniqueness of solutions to
  characteristic evolution in Bondi-Sachs coordinates in General Relativity}},
  Phys.\ Rev.\ D \textbf{59} (1999), 084012.

\bibitem{GerochWinicour81}
R.P. Geroch and J.~Winicour, \emph{Linkages in general relativity}, Jour.\
  Math.\ Phys. \textbf{22} (1981), 803--812. \MR{617326}

\bibitem{HintzdSBH}
Peter Hintz, \emph{Black hole gluing in de {S}itter space}, Commun.\ PDEs
  \textbf{46} (2021), 1280--1318, arXiv:2001.10401 [math.AP]. \MR{4279966}

\bibitem{HIW}
S.~Hollands, A.~Ishibashi, and R.M. Wald, \emph{A higher dimensional stationary
  rotating black hole must be axisymmetric}, Commun.\ Math.\ Phys. \textbf{271}
  (2007), 699--722, arXiv:gr-qc/0605106.

\bibitem{Israel66}
W.~Israel, \emph{Singular hypersurfaces and thin shells in general relativity},
  Il Nuovo Cimento \textbf{44B} (1966), 1--14.

\bibitem{Kehle:2022uvc}
C.~Kehle and R.~Unger, \emph{{Gravitational collapse to extremal black holes
  and the third law of black hole thermodynamics}},  (2022), arXiv:2211.15742
  [gr-qc].

\bibitem{KhanPenrose}
K.~A. Khan and R.~Penrose, \emph{{Scattering of two impulsive gravitational
  plane waves}}, Nature \textbf{229} (1971), 185--186.

\bibitem{KorbiczTafel}
J.~Korbicz and J.~Tafel, \emph{Lagrangian and {H}amiltonian for the
  {B}ondi-{S}achs metrics}, Class.\ Quantum Grav. \textbf{21} (2004),
  3301--3308. \MR{MR2072137 (2005g:83012)}

\bibitem{Luk}
J.~Luk, \emph{On the local existence for the characteristic initial value
  problem in general relativity}, Int.\ Math.\ Res.\ Not.\ IMRN (2012),
  4625--4678. \MR{2989616}

\bibitem{MaedlerWinicour}
T.~M{\"a}dler and J.~Winicour, \emph{{Bondi-Sachs formalism}}, Scholarpedia
  \textbf{11} (2016), 33528, arXiv:1609.01731 [gr-qc].

\bibitem{MantoulidisSchoen}
Christos Mantoulidis and R.~Schoen, \emph{On the {B}artnik mass of apparent
  horizons}, Class. Quantum Grav. \textbf{32} (2015), 205002, 16. \MR{3406373}

\bibitem{MarsCharacteristic2}
M.~Mars and G.~S\'anchez-P\'erez, \emph{{Covariant definition of Double Null
  Data and geometric uniqueness of the characteristic initial value problem}},
  Jour. Phys. A (2023), 255203, arXiv:2301.02722 [gr-qc].

\bibitem{Moncrief75}
V.~Moncrief, \emph{Spacetime symmetries and linearization stability of the
  {E}instein equations {I}}, Jour.\ Math.\ Phys. \textbf{16} (1975), 493--498.

\bibitem{VinceJimcompactCauchyCMP}
V.~Moncrief and J.~Isenberg, \emph{Symmetries of cosmological {C}auchy
  horizons}, Commun.\ Math.\ Phys. \textbf{89} (1983), 387--413. \MR{709474
  (85c:83026)}

\bibitem{RendallCIVP}
A.D. Rendall, \emph{Reduction of the characteristic initial value problem to
  the {C}auchy problem and its applications to the {E}instein equations},
  Proc.\ Roy.\ Soc.\ London A \textbf{427} (1990), 221--239. \MR{MR1032984
  (91a:83004)}

\bibitem{RodnianskiShlapentokh}
I.~Rodnianski and Y.~Shlapentokh-Rothman, \emph{The asymptotically self-similar
  regime for the {E}instein vacuum equations}, Geom.\ Funct.\ Anal. \textbf{28}
  (2018), 755--878. \MR{3816523}

\bibitem{Sachs}
R.K. Sachs, \emph{Gravitational waves in general relativity {VIII.} {Waves} in
  asymptotically flat spacetime}, Proc.\ Roy.\ Soc.\ London A \textbf{270}
  (1962), 103--126. \MR{MR0149908 (26 \#7393)}

\bibitem{SmithWeinstein2}
B.~Smith and G.~Weinstein, \emph{On the connectedness of the space of initial
  data for the {E}instein equations}, Electron.\ Res.\ Announc.\ Amer.\ Math.\
  Soc. \textbf{6} (2000), 52--63. \MR{1777856}

\bibitem{SmithWeinstein1}
\bysame, \emph{Quasiconvex foliations and asymptotically flat metrics of
  non-negative scalar curvature}, Commun.\ Anal.\ Geom. \textbf{12} (2004),
  511--551. \MR{2128602}

\end{thebibliography}

\end{document}